# FINEX: A Fast Index for Exact & Flexible Density-Based Clustering (Extended Version with Proofs)*

KONSTANTIN EMIL THIEL, University of Salzburg, Austria

DANIEL KOCHER, University of Salzburg, Austria

NIKOLAUS AUGSTEN, University of Salzburg, Austria

THOMAS HÜTTER, University of Salzburg, Austria

WILLI MANN, Celonis SE, Germany

DANIEL ULRICH SCHMITT, University of Salzburg, Austria

Density-based clustering aims to find groups of similar objects (i.e., clusters) in a given dataset. Applications include, e.g., process mining and anomaly detection. It comes with two user parameters ($\varepsilon$, $MinPts$) that determine the clustering result, but are typically unknown in advance. Thus, users need to interactively test various settings until satisfying clusterings are found. However, existing solutions suffer from the following limitations: (a) *Ineffective pruning* of expensive neighborhood computations. (b) *Approximate* clustering, where objects are falsely labeled noise. (c) *Restricted* parameter tuning that is limited to $\varepsilon$ whereas $MinPts$ is constant, which reduces the explorable clusterings. (d) *Inflexibility* in terms of applicable data types and distance functions.

We propose FINEX, a linear-space index that overcomes these limitations. Our index provides *exact* clusterings and can be queried with either of the two parameters. FINEX avoids neighborhood computations where possible and reduces the complexities of the remaining computations by leveraging fundamental properties of density-based clusters. Hence, our solution is efficient and flexible regarding data types and distance functions. Moreover, FINEX respects the original and straightforward notion of density-based clustering. In our experiments on 12 large real-world datasets from various domains, FINEX frequently outperforms state-of-the-art techniques for exact clustering by orders of magnitude.

CCS Concepts: • **Information systems → Clustering**; **Clustering and classification**.

Additional Key Words and Phrases: density-based clustering, index, OPTICS, DBSCAN

## 1 INTRODUCTION

Clustering aims to identify groups of similar objects (i.e., clusters) in datasets. It is a vital task, as it helps to explore and understand dataset characteristics [11]. In density-based clustering, these clusters are high-density regions of objects separated by regions of low density. The popular DBSCAN [8] algorithm computes this type of clustering and is successfully deployed in various fields, such as astronomy [19] or anomaly detection [23]. In particular, our research is motivated by the process mining domain, where DBSCAN clusters are used to analyze large amounts of business processes modeled as sets [15]. DBSCAN has two user parameters ($\varepsilon$, $MinPts$) that define the minimum density of clusters: Objects with at least $MinPts$ neighbors within a radius of $\varepsilon$ are classified as *cores* that spawn or expand clusters at their locations. If an object with fewer neighbors is located farther than $\varepsilon$ from core objects, it is considered noise.

***DBSCAN's Limitation.*** Selecting ($\varepsilon$, $MinPts$) is difficult due to its heavy influence on the clustering result. Sometimes, there exists no unique setting that captures all meaningful clusters: Consider Figure 1, which shows two density-based clusterings w.r.t. different parameters. The top-right cluster (green) in Figure 1a requires a lower density than the bottom-right clusters (yellow and purple) in Figure 1b. If no prior knowledge is available, users have to recompute DBSCAN with different settings until they find a suitable clustering. In the era of big data, this poses a

---

severe problem to DBSCAN due to its inefficiency. Without neighborhood index, the algorithm involves $O(n^2)$ distance computations, which renders *interactive* clustering infeasible for large datasets. For example, even with a state-of-the-art neighborhood index, clustering the CELONIS-1 dataset (cf. Section 6) takes up to 9 hours in our experiments, which is prohibitively long for process mining companies dealing with thousands of clustering queries per day.

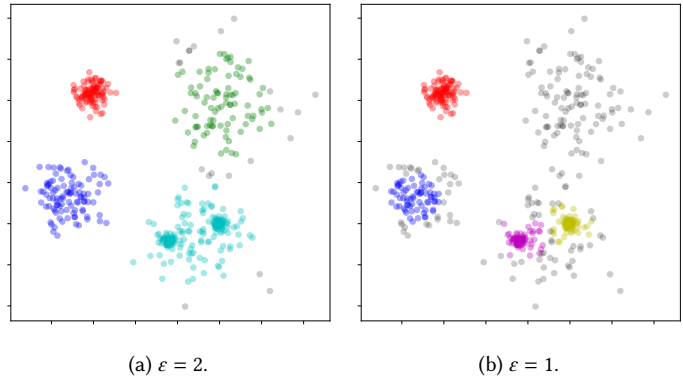

(a) $\varepsilon = 2$.      (b) $\varepsilon = 1$.

Fig. 1. Density-based clustering for two different $\varepsilon$ values in 2-d space. Both clusterings might be meaningful.

***State-of-the-Art Index.*** The OPTICS technique [2] attempts to mitigate the problem of choosing a distance threshold in advance: It computes a *cluster ordering*, which can be used as an index that returns clusterings when queried with some threshold $\varepsilon^* \leq \varepsilon$. However, OPTICS only provides *approximate* clusterings as it falsely labels some cluster objects as noise. This is a crucial issue in scenarios where *exact* results are required to correctly distinguish noise from clusters, e.g., in anomaly detection [23]. Moreover, OPTICS does not allow to tune $MinPts$, which limits its applicability.

***FINEX.*** We propose FINEX, a **f**ast linear-space **in**dex for **ex**act and flexible density-based clustering, which is applicable to any data type and distance function. It overcomes DBSCAN's complexity problem and OPTICS' inaccuracy without introducing additional limitations or overhead. This sets FINEX apart from other approaches that either compromise the original notion of density-based clustering [6], only work in specific settings [9], or require additional user parameters [16].

***Contributions.*** Our contributions are destined to enable interactive density-based clusterings of large datasets:

(1) *Exactness & Parameter Tuning:* FINEX supports two new query types: $\varepsilon^*$-queries and $MinPts^*$-queries, which return *exact* clusterings. Once FINEX has been built for an generating $(\varepsilon, MinPts)$ pair, users can query *all* clusterings w.r.t. either $(\varepsilon^*, MinPts)$ with $\varepsilon^* \leq \varepsilon$ or $(\varepsilon, MinPts^*)$ with $MinPts^* \geq MinPts$. Unlike OPTICS, FINEX is able to correctly identify *all* cluster objects.

(2) *Efficiency & Flexibility:* In contrast to DBSCAN, FINEX answers all queries fast and without recomputation from scratch. Its efficiency is based on the fact that dense clusters are subsets of sparse clusters, a result that holds for any data type and distance function. FINEX leverages this fact to reduce the number and cost of neighborhood computations, and relies only on standard requirements for density-based clustering (i.e., a symmetric distance function for pairs of objects).

(3) *Fast Linear-Space Index:* FINEX is designed as an in-memory data structure providing a linear space complexity in the number of objects. Precisely, our index is a permutation of the dataset where each object is equipped with two special distance measures, a density count and a reference to its densest neighbor. FINEX leverages these attributes to identify cluster objects that would be falsely labeled as noise and to find connected components if a sparse cluster decomposes into multiple dense clusters.

(4) *Linear-Time Clustering:* As with OPTICS, FINEX features linear-time approximate clustering for $\varepsilon^* \leq \varepsilon$. Thereby, FINEX returns clusterings at least as accurate as OPTICS' clusterings. Moreover, unlike OPTICS, we provide an *exact* clustering in linear time if the query parameters are identical to the generating $(\varepsilon, MinPts)$ pair, for which FINEX was constructed.

(5) *Experimental Evaluation:* We complement our theoretical results with an experimental evaluation on 12 large real-world datasets that either contain set-valued or multi-dimensional vector objects from various domains using the Jaccard and Euclidean distance. We compare the clustering runtime of FINEX against two competitors that compute exact clusterings from scratch: a highly optimized DBSCAN and the state-of-the art algorithm AnyDBC [16]. Moreover, we compare the index building time and the accuracy of FINEX and OPTICS. Our results suggest that FINEX outperforms its competitors in almost every scenario, often by orders of magnitude.

***Paper Outline.*** The remainder of this paper is organized as follows: Section 2 provides an overview of related literature and Section 3 introduces the most important definitions and discusses limitations of existing solutions. In Section 4, we rigorously formalize OPTICS and study fundamental properties of cluster orderings to set the stage for our solution. FINEX and its theoretical guarantees are then introduced in Section 5. To support our theoretical findings, we experimentally evaluate FINEX against state-of-the-art solutions in Section 6. We finally draw conclusions in Section 7. Moreover, we provide proofs of all our theoretical results in Appendix A.

## 2 RELATED WORK

To the best of our knowledge, FINEX introduces the *first* linear-space data structure efficiently providing *exact* density-based clusterings. While DBSCAN [8] also generates exact clusterings, it has to be re-computed from scratch whenever an input parameter is changed. A formalization of exactness is introduced in Definition 3.5, following the original notion of density-based clustering.

***Indexing Structure.*** The OPTICS algorithm [2] deploys a linear *cluster ordering* to provide dense clusters. In fact, FINEX and OPTICS leverage the same fundamental property: dense clusters are subsets of sparse ones, which can be encoded in the ordering. However, OPTICS does not fully exploit this property, and thus, cannot report exact clusterings. FINEX provides this feature and further allows for tuning the *MinPts* parameter, expanding the explorable solution space. Moreover, FINEX' linear-time approximate clustering produces more accurate results than OPTICS.

Developing FINEX requires a thorough formalization of cluster orderings. Achtert et al. [1] were the first to formalize OPTICS in their work to implement updates to a precomputed OPTICS-ordering. Although their formalization cannot be directly used for our purposes, their basic ideas inspired our formal approach.

Gan and Tao [9] developed an efficient approximate version of OPTICS that works only for vector data and the $L_p$ norm based on a different formalization. In contrast, FINEX produces exact results and is flexible in terms of data types and distance functions.

***Efficient Neighborhood Computations.*** A natural step towards interactive density-based clustering is to speed-up $\varepsilon$-neighborhood computations that dominate DBSCAN's runtime. There exist data structures for efficient neighborhood computations, e.g., R-trees [10] or kd-trees [4] for vector data, and M-trees [7] for metric spaces. Mann et al. [18] compare several state-of-the-art techniques for set data. Since FINEX supports the same data types and distance functions as DBSCAN, it can be combined with any of these techniques.

***Pruning Neighborhood Computations***. FINEX' efficiency is attributed to reducing the number of neighborhood computations compared to DBSCAN, which computes neighborhoods for all objects of a dataset. The AnyDBC [16, 17] algorithm also performs neighborhood computations only for subsets of objects. These objects spawn partial clusters that are successively connected until an exact clustering is reached. Compared to FINEX, AnyDBC needs additional user parameters and is limited to metric distance functions (for which the triangle inequality must hold).

***Filtering Approach***. Brecheisen et al. [5] discovered that clusters computed with the true distance function are subsets of clusters computed with a lower-bounding distance function. Thus, a clustering itself can be used as a filter for another clustering. However, Brecheisen et al. use this result to compute one specific clustering while FINEX retrieves an arbitrary number of clusterings.

***Other Approaches***. Jang and Jiang [13] suggest to speed-up density-based clustering by drawing random samples of objects for which neighborhoods are computed. Compared to FINEX, this approach yields an approximation of the exact clustering and requires additional assumptions about the statistical data distribution.

Campello et al. [6] propose an alternative notion of density-based clustering, in which clusters are only composed of *core objects*. While this notion avoids OPTICS' approximation problem as OPTICS is exact w.r.t. core objects (cf. Section 4), their approach remains an approximation w.r.t. the original notion of density-based clustering. FINEX provides an exact solution in both scenarios.

## 3 PRELIMINARIES

In this section, we recap density-based clustering and its most important definitions. Moreover, we introduce a novel definition of the *exact clustering* computed by DBSCAN. Subsequently, we explain the OPTICS algorithm and the two quantities, *core distance* and *reachability distance*, central for the indexing structure it computes. We conclude this section with an illustration of OPTICS' limitations that we overcome in this paper.

### 3.1 Density-Based Clustering

Density-based clustering is flexible in terms of supported data types and distance functions. It only requires a dataset $D$ and a symmetric distance function $d : D \times D \rightarrow \mathbb{R}_{\geq 0}$. The density is defined by two parameters: (1) A *distance threshold* $\varepsilon \in \mathbb{R}_{\geq 0}$ that defines the $\varepsilon$-neighborhood $N_\varepsilon(p)$ of an object $p$ in a dataset $D$:

$$N_\varepsilon(p) = \{q \in D \,:\, d(p, q) \leq \varepsilon\}$$

(2) A *size threshold*, called *MinPts* $\in \mathbb{N}$, that defines the *core* status of objects: An object with at least *MinPts* objects in its $\varepsilon$-neighborhood is called *core object*. Objects with $\varepsilon$-neighborhoods smaller than *MinPts* that are located in the $\varepsilon$-neighborhood of another core object are called *border objects* of a cluster. Note that $p \in N_\varepsilon(p)$ always holds. The relationship between clusters, core objects, and border objects is expressed in Definitions 3.1–3.4.

*Definition 3.1 (Directly Density-Reachable [8]).* Let $p, q \in D$. $p$ is *directly density-reachable* from $q$ w.r.t. $(\varepsilon, MinPts)$ if

(1) $|N_\varepsilon(q)| \geq MinPts$
(2) $p \in N_\varepsilon(q)$

An object can only be directly density-reachable from a core object, which implies that this relation is not symmetric. Natural extensions of this concept are given by the following definitions.

*Definition 3.2 (Density-Reachable [8]).* Let $p, q \in D$. $p$ is *density-reachable* from $q$ w.r.t. ($\varepsilon$, *MinPts*) if $\exists p_1, \ldots, p_n \in D$ such that

(1) $p_1 = q, p_n = p$
(2) $p_{i+1}$ is directly density-reachable from $p_i$ w.r.t. ($\varepsilon$, *MinPts*) for all $i = 1, \ldots, n - 1$

*Definition 3.3 (Density-Connected [8]).* Let $p, q \in D$. $p$ is *density-connected* to $q$ w.r.t. ($\varepsilon$, *MinPts*) if $\exists o \in D$ such that both $p$ and $q$ are density-reachable from $o$ w.r.t. ($\varepsilon$, *MinPts*).

Density-reachable is a transitive relation and density-connected is additionally symmetric. With these concepts in mind, we are ready to introduce the formal definition of density-based clusters.

*Definition 3.4 (Cluster and Noise [8]).* Any non-empty subset $K \subseteq D$ satisfying the following conditions is a *cluster* w.r.t. ($\varepsilon$, *MinPts*):

(1) Maximality: $\forall p, q \in D$ : if $p \in K$ and $q$ is density-reachable from $p$ w.r.t. ($\varepsilon$, *MinPts*), then also $q \in K$.
(2) Connectivity: $\forall p, q \in K$ : $p$ is density-connected to $q$ w.r.t. ($\varepsilon$, *MinPts*).

Every object that is not contained in any cluster is considered *noise*.

Note that there might exist border objects that are density-reachable from core objects of *different* clusters. We refer to such border objects as *ambiguous* since they belong to more than one cluster, according to Definition 3.4.

**Exactness of DBSCAN.** The DBSCAN algorithm computes a density-based clustering for a dataset $D$ by detecting an unprocessed core object $o \in D$ and creating a new cluster with all objects that are density-reachable from o w.r.t. ($\varepsilon$, *MinPts*). DBSCAN terminates when all objects have been processed, and hence, all clusters have been computed. Note that the algorithm assigns all density-connected objects to exactly one cluster. (Ambiguous) border objects are assigned to the cluster defined by the core object from which they are detected first. Therefore, DBSCAN's result is what we refer to as an *exact clustering* in our new Definition 3.5.

*Definition 3.5 (Exact Clustering).* Let $K_1, \ldots, K_n$ be the density-based clusters of $D$ w.r.t. ($\varepsilon$, *MinPts*) according to Definition 3.4. An *exact clustering* is a partitioning $E_1, \ldots, E_n$ of $D$ such that

(1) $E_i \subseteq K_i$
(2) all core objects of $K_i$ are in $E_i$
(3) every border object is in exactly one partition that corresponds to a cluster it belongs to.

In other words, the difference between an exact clustering and a density-based clustering is that the exact clustering assigns each ambiguous border object to exactly one cluster.

### 3.2 State-of-the-Art Indexing Technique

The OPTICS algorithm [2] attempts to make density-based clustering more interactive. The idea is that one computation of OPTICS substitutes several runs of DBSCAN. As with DBSCAN, users choose parameters $\varepsilon \in \mathbb{R}_{\geq 0}$ and *MinPts* $\in \mathbb{N}$. OPTICS then computes a *cluster ordering* that contains information about any density-based clustering w.r.t. ($\varepsilon^*$, *MinPts*) such that $0 \leq \varepsilon^* \leq \varepsilon$. However, OPTICS-orderings also miss substantial information about these clusterings (cf. Section 3.3). To understand this, we first introduce the ordering's constituent parts in Definitions 3.6−3.8.

*Definition 3.6 (MinPts-Distance of $p \in D$ [2]).*

$$M(p) = \min\{\delta \in \mathbb{R}_{\geq 0} : |N_\delta(p)| \geq MinPts\}$$

*Definition 3.7 (Core Distance of $p \in D$ [2]).*

$$C_{\varepsilon, MinPts}(p) = \begin{cases} M(p), & \text{if } |N_\varepsilon(p)| \geq MinPts \\ \infty, & \text{otherwise} \end{cases}$$

Consider any core object $p \in D$ and let $\delta < C_{\varepsilon, MinPts}(p)$. It follows that $|N_\delta(p)| < MinPts$. Thus, the core distance of $p$ is the smallest distance such that $p$ is still a core object. Noise and border objects always have a core distance of $\infty$.

*Definition 3.8 (Reachability Distance of $q \in D$ w.r.t. $p \in D$ [2]).*

$$R_{\varepsilon, MinPts}(q, p) = \begin{cases} \max\{C_{\varepsilon, MinPts}(p), d(p, q)\}, & \text{if } |N_\varepsilon(p)| \geq \text{MinPts} \\ \infty, & \text{otherwise} \end{cases}$$

The definition of the reachability distance is related to the concept of density-reachability: let $p \in D$ be a core object and $q \in N_\varepsilon(p)$, then $R_{\varepsilon, MinPts}(q, p)$ is the smallest distance threshold such that $q$ is still directly density-reachable from $p$. At any smaller distance, either $q$ would be outside $p$'s neighborhood or $p$ would lose its core property, meaning that no object is directly density-reachable from $p$. Note that, in general, $R_{\varepsilon, MinPts}(q, p) \neq R_{\varepsilon, MinPts}(p, q)$. We illustrate the distances in Figure 2.

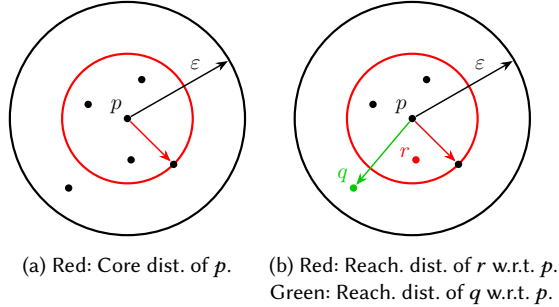

(a) Red: Core dist. of $p$.   (b) Red: Reach. dist. of $r$ w.r.t. $p$.
Green: Reach. dist. of $q$ w.r.t. $p$.

Fig. 2. Core and Reachability distances (cf. Definitions 3.7 and 3.8) exemplified for $MinPts = 5$.

***OPTICS' Trick.*** OPTICS uses the fact that dense clusters are completely contained in sparse ones. However, Ankerst et al. [2] only provide an intuition of this property, a formal proof is missing. We complement their work with Proposition 3.9. The proof is provided in Appendix A.

PROPOSITION 3.9 ($\varepsilon$-NESTED CLUSTERS). *Let $0 \leq \varepsilon^* \leq \varepsilon$. For every cluster $K^* \subseteq D$ w.r.t. ($\varepsilon^*$, MinPts) there is a cluster $K \subseteq D$ w.r.t. ($\varepsilon$, MinPts) such that $K^* \subseteq K$.*

OPTICS attempts to exploit Proposition 3.9 by establishing a permutation of the objects in $D$, in which dense regions are surrounded by sparse ones. Thereby, the algorithm uses reachability distance as a measure of density, with high density corresponding to a low reachability distance.

**Algorithm**. OPTICS can be implemented as a nested loop as follows: In the outer loop, an arbitrary object $o$ is drawn from $D$ and $o$ is appended to an initially empty cluster ordering $O$. If $o$ is a core object w.r.t. ($\varepsilon$, $MinPts$), $o$'s neighbors are inserted into a priority queue $Q$ and the control switches to an inner loop that successively pops objects from $Q$ until it is empty again. The priority of objects in $Q$ is initially determined by their reachability distance w.r.t. $o$. The object $p$ with currently smallest reachability distance is always popped first (high priority) and appended to the ordering $O$. If $p$ is a core object w.r.t. ($\varepsilon$, $MinPts$), $p$'s neighbors are added to $Q$ with their reachability distance computed w.r.t. $p$. If a neighbor is already contained in $Q$, its current reachability distance is compared to the new reachability distance w.r.t. $p$. If the new one w.r.t. $p$ is smaller than the current one, the neighbor's priority is updated accordingly.

OPTICS processes each object exactly once (i.e., objects that are already appended to $O$ won't be reinserted into $Q$). Whenever an object is appended to $O$, also its core distance and the reachability distance at which it was popped out of $Q$ are recorded. Objects processed in the outer loop are assigned a reachability distance of $\infty$. Hence, the recorded reachability distance of every processed object $x \in O$ is the smallest distance threshold at which $x$ is density-reachable from any object was appended to $O$ before.

**Clustering**. When the algorithm terminates, Proposition 3.9 is reflected in the ordering $O$: dense regions of $D$ are surrounded by the sparse regions from which they are reachable. As a result, any clustering w.r.t. $\varepsilon^* \leq \varepsilon$ and a fixed $MinPts$ can be queried from $O$. Querying is best illustrated in a so-called *reachability plot*, which plots reachability distances of objects in processing order from left to right. Figure 3 shows this for the dataset from Figure 1. A clustering can be extracted by cutting the vertical axis of the plot at $\varepsilon^*$ and assigning the objects in each *valley* to a separate cluster. Moreover, for each valley, an initial core object $o$, which is located directly before the valley and identified via $C_{\varepsilon, MinPts}(o) \leq \varepsilon^*$, is added to the respective cluster.

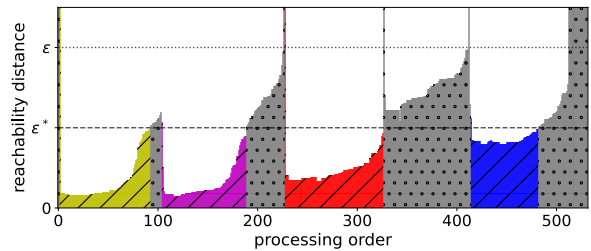

Fig. 3. Reachability plot with OPTICS-clustering of the dataset from Figure 1 w.r.t. $\varepsilon^* = \varepsilon/2$.

### 3.3 Limitations

Both DBSCAN and OPTICS suffer from serious limitations, which are outlined in this section.

**Efficiency**. To provide a clustering for a given dataset, DBSCAN has to compute the $\varepsilon$-neighborhoods of all objects in the dataset. It then assesses the core status of these objects and determines the connected components among the detected core objects. A single neighborhood computation requires up to $O(n)$ distance computations (where $n$ denotes dataset size). As a result, DBSCAN comes with $O(n^2)$ complexity, which renders interactive clustering infeasible when working with large datasets.

Table 1. Core objects w.r.t. $\varepsilon$ and $MinPts = 4$ from Figure 4 (Euclidean distances relative to $\varepsilon$).

| Object | Core Dist. | Sorted $\varepsilon$-neighborhood |
|--------|-----------|-----------------------------------|
| C | 1 | (A, $\sqrt{5}/4$), (D, $1/\sqrt{2}$), (B, 1), (E, 1) |
| D | 3/4 | (C, $1/\sqrt{2}$), (E, $1/\sqrt{2}$), (A, 3/4), (F, 1) |
| H | $1/\sqrt{2}$ | (G, $\sqrt{5}/4$), (J, $\sqrt{5}/4$), (I, $1/\sqrt{2}$), (K, 1) |
| I | 3/4 | (H, $1/\sqrt{2}$), (K, $1/\sqrt{2}$), (F, 3/4), (J, 3/4) |
| J | 3/4 | (H, $\sqrt{5}/4$), (K, $\sqrt{5}/4$), (I, 3/4), (G, 1) |
| K | 1 | (J, $\sqrt{5}/4$), (I, $1/\sqrt{2}$), (H, 1) |

***Parameter Tuning & Approximation.*** Even though an OPTICS-ordering can be queried in $O(n)$ time to obtain a clustering, we identify two major problems: (i) The *MinPts* parameter cannot be changed, which limits the clustering solutions that can be explored from an OPTICS-ordering. (ii) Clusterings queried from an OPTICS-ordering w.r.t. $\varepsilon^* \leq \varepsilon$ are only approximations of the exact clusterings (cf. Definition 3.5) produced by DBSCAN. We show the severity of (ii) in Example 3.10.

*Example 3.10.* Let $\varepsilon^* = 3\varepsilon/4$ and $MinPts = 4$. Figure 4 contains a small 2-d dataset, whose objects are colored according an exact clustering w.r.t. ($\varepsilon^*$, *MinPts*). There are clusters $K_1 = \{A, C, D, E\}$ (blue) and $K_2 = \{F, G, H, I, J, K\}$ (yellow), and the noise group $N = \{B\}$ (black). Figure 4b shows the reachability plot of an OPTICS-ordering computed w.r.t. ($\varepsilon$, *MinPts*) with bars colored according to the clustering queried from the OPTICS-ordering. **We observe:** OPTICS' clusters are substantially smaller than the exact ones, as $K_1$ misses 50% of its objects and $K_2$ misses a third of its objects, respectively. All missing objects are labeled noise.

For the sake of reproducibility of Example 3.10, we include $\varepsilon$-neighborhoods of core objects in this example in Table 1. Thus, interested readers can reconstruct the OPTICS-ordering and its clustering following the description in Section 3.2 or the algorithms in Ankerst et al. [2]. It is worth mentioning that Example 3.10 does not depict the worst case. Indeed, clustering based on OPTICS could miss *every single border object*, which is not the case here.

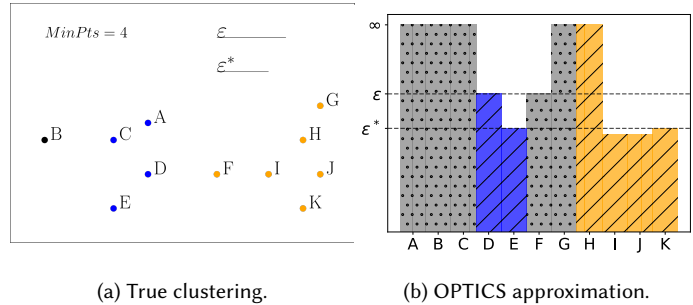

(a) True clustering.　　　　(b) OPTICS approximation.

Fig. 4. Clustering w.r.t. $\varepsilon^* = 3\varepsilon/4$ and $MinPts = 4$.

## 4 FORMALIZATION OF OPTICS

We use OPTICS as the starting point for developing a new index, FINEX, that overcomes the limitations outlined in Section 3.3. Thus, we thoroughly study and formalize the OPTICS-ordering and discuss the properties of objects falsely classified as noise by OPTICS.

---

**Algorithm 1:** QueryClustering($O, \varepsilon^*$) [2]

---

**Input:** cluster ordering $O$, distance threshold $\varepsilon^*$
**Output:** set of clusters and noise group

1   $Noise \leftarrow \emptyset$
2   $Clustering \leftarrow \emptyset$
3   $S \leftarrow \emptyset$                                             `// S collects cluster objects`
4   **foreach** $x \in O$ **do**
5     **if** $x.R > \varepsilon^*$ **then**
6        **if** $x.C \leq \varepsilon^*$ **then**
7           **if** $S \neq \emptyset$ **then**
8              $Clustering \leftarrow Clustering \cup \{S\}$
9              $S \leftarrow \emptyset$                           `// begin new cluster`
10          $S \leftarrow S \cup \{x\}$
11        **else**
12           $Noise \leftarrow Noise \cup \{x\}$
13     **else**
14        $S \leftarrow S \cup \{x\}$
15   **if** $S \neq \emptyset$ **then**
16     $Clustering \leftarrow Clustering \cup \{S\}$
17   $Clustering \leftarrow Clustering \cup \{Noise\}$
18   **return** $Clustering$

---

***Cluster Ordering***. From Section 3.2, we know that OPTICS computes a permutation of the input dataset $D$, augmented with core and reachability distances for each $x \in D$. We introduce a precise formalization in Definition 4.1, which is inspired by Achtert et al. [1].

*Definition 4.1 (OPTICS-Ordering).* Let $D$ be a dataset and $\min \emptyset = \infty$. An *OPTICS-ordering* $O_{\varepsilon, MinPts}$ is a permutation of $D$ where each object $x \in D$ is equipped with the following attributes:

    $x.P$: Permutation number in $\{1, \ldots, |D|\}$

    $x.C$: Core distance, $C_{\varepsilon, MinPts}(x)$

    $x.R$: Reachability distance, $\min\limits_{p \in D, p.P < x.P} \{R_{\varepsilon, MinPts}(x, p)\}$

Note that Definition 4.1 uses an attribute $x.P$ to implement the permutation, which will simplify our subsequent theoretical considerations. However, in practice, this attribute can be avoided by aligning the objects according to their processing order. The OPTICS algorithm [2] computes an ordering according to Definition 4.1 to report approximate clusterings w.r.t. $\varepsilon^* \leq \varepsilon$ and *MinPts*.

***Cluster Extraction***. We investigate the clusters queried from an OPTICS-ordering, and therefore, briefly introduce the original OPTICS querying technique by Ankerst et al. [2] in a version that complies with the notation in the remainder of this paper. Its pseudocode is available in Algorithm 1. The query technique basically traverses all objects $x \in O_{\varepsilon, MinPts}$ in processing order and uses their core- and reachability distance attributes ($x.C$ and $x.R$) to assign them to the current cluster $S$ or to the noise group, respectively. Note that a new cluster is started once a core object $x$ w.r.t. ($\varepsilon^*$, *MinPts*) is found for which $x.R > \varepsilon^*$ (cf. line 6). The cluster is then expanded until the next object with a reachability distance greater than $\varepsilon^*$ is encountered (cf. line 5). Recall that Algorithm 1 can be intuitively imagined as cutting the vertical axis of a reachability plot at $\varepsilon^*$ and adding an initial core object to the valleys that are formed (cf. Figure 3).

***Cluster Properties***. We next list the precise properties of the elements in the cluster set $S$ from Algorithm 1 in Definition 4.2. Note that we refer to $S$ as an *approximate cluster* since, as we will subsequently show, $S$ only approximates a density-based cluster.

*Definition 4.2 (Approximate Cluster).* We refer to a subset $S = \{x_1, \ldots, x_n\} \subseteq D$ as an *approximate cluster* w.r.t. $\varepsilon^* \leq \varepsilon$ and *MinPts* if its elements meet the following conditions:

(1) $x_{i+1}.P = x_i.P + 1$ for all $i = 1, \ldots, n-1$
(2) $x_1.C \leq \varepsilon^*$ and $x_1.R > \varepsilon^*$
(3) $x_i.R \leq \varepsilon^*$ for all $i = 2, \ldots, n$
(4) Either $x_n.P = |D|$ or it holds for $y \in D$ with $y.P = x_n.P + 1$ that $y.R > \varepsilon^*$

The properties listed in Definition 4.2 can be expressed as follows: (1) $S$ consists of objects that are adjacent in $O_{\varepsilon, MinPts}$. (2) $x_1$ is a core object w.r.t. ($\varepsilon^*$, *MinPts*) and $x_1$ is not density-reachable w.r.t. ($\varepsilon^*$, *MinPts*) from any other object processed before. (3) each object $x_i$, $i = 2, \ldots, n$ is density-reachable w.r.t. ($\varepsilon^*$, *MinPts*) from some other objects with lower permutation number. (4) $x_n$ is either the last object in $O_{\varepsilon, MinPts}$ or the object $y$, processed immediately after $x_n$, is not density-reachable w.r.t. ($\varepsilon^*$, *MinPts*) from another object with lower permutation number.

***Missing Objects***. OPTICS processes each object $x \in D$ only once, and hence, cannot globally minimize the reachability distance attribute $x.R$, which is used by Algorithm 1 to determine $x$' cluster membership. For instance, if $x$ is a border object w.r.t. ($\varepsilon$, *MinPts*) that is processed in the outer OPTICS loop, then $x.R = \infty$ is assigned (cf. Section 3). Consequently, $x$ will *always* be falsely labeled as noise by Algorithm 1, even if $\varepsilon^* = \varepsilon$. In Theorem 4.3, we precisely state how an approximate cluster $S \subseteq O_{\varepsilon, MinPts}$ computed by OPTICS relates to a density-based cluster according to Definition 3.4. See Appendix A for a proof.

THEOREM 4.3 (APPROXIMATE OPTICS CLUSTERS). *Let $O_{\varepsilon, MinPts}$ be a cluster ordering computed by OPTICS. We consider an approximate cluster $S$ w.r.t. ($\varepsilon^*$, MinPts) according to Definition 4.2. The following propositions hold:*

(a) $\exists$ *a density-based cluster $K$ w.r.t. $\varepsilon^*$ and MinPts according to Definition 3.4 such that $S \subseteq K$.*
(b) $\forall p \in K : p.P \leq x_n.P$
(c) $\forall p \in K : |N_{\varepsilon^*}(p)| \geq MinPts \implies p \in S$

Theorem 4.3 can be read as an instruction to turn an approximate clustering by OPTICS into an exact one: It states that each approximate cluster $S$ is a subset of some density-based cluster $K$. Moreover, all objects in $K \setminus S$ are processed before $x_1 \in S$. Finally, $S$ contains all core objects of $K$, and thus, $K \setminus S$ exclusively consists of border objects. In other words, the search space for missing border objects may consist of all objects processed before $x_1$ that do not belong to any other approximate cluster (and hence, are labeled as noise). A naive way to turn an approximate clustering by OPTICS into an exact one according to Definition 3.5 is to check for each object in the search space whether it is in fact located in the $\varepsilon^*$-neighborhood of some core object. However, depending on the amount of noise, this can be computationally expensive. Therefore, the rationale of our own index will be to limit the search space as far as possible, and thus, reduce the effort at query time.

## 5 CLUSTERING WITH FINEX

In this section, we introduce our novel linear-space index called FINEX, which enables a fast extraction of an *exact* density-based clustering for changing threshold parameters, i.e., either $\varepsilon^* \leq \varepsilon$ or $MinPts^* \geq MinPts$. We present the construction of FINEX for a given pair $(\varepsilon, MinPts)$ and discuss how FINEX can be used to extract an approximate clustering that is more accurate than the OPTICS result. We also show that FINEX returns an exact clustering if $\varepsilon^* = \varepsilon$ and $MinPts^* = MinPts$ (unlike OPTICS). Finally, we discuss how to use FINEX to efficiently derive an exact clustering w.r.t. either $\varepsilon^* < \varepsilon$ or $MinPts^* > MinPts$, respectively. The proofs for all theory in this section can be found in Appendix A.

### 5.1 Constructing FINEX

This section summarizes the FINEX index structure and how to construct it for a given dataset $D$ and a generating pair $(\varepsilon, MinPts)$.

*Index Structure.* Our ultimate goal of extracting exact density-based clusterings is also reflected in the structure and the interface of FINEX. Essentially, FINEX is a permutation of the input dataset $D$ and each object $x \in D$ is augmented with a quintuple $(P, C, R, N, F)$, where $x.P$ and $x.N$ are integers, $x.C$ and $x.R$ are floating-point numbers, and $x.F$ is a reference (cf. Definition 5.1). Consequently, FINEX requires linear space in the dataset size $|D|$.

Our index takes two parameters, a dataset $D$ and a generating pair $(\varepsilon, MinPts)$, and returns a *FINEX-ordering* as output. A FINEX-ordering enhances the OPTICS-ordering and is defined as follows.

*Definition 5.1 (FINEX-Ordering).* Let $D$ be a dataset. A *FINEX-ordering* $\widetilde{O}_{\varepsilon, MinPts}$ is a permutation of $D$ where each object $x \in D$ is equipped with the following attributes:

$x.P$: Permutation number, $P \in \{1, \ldots, |D|\}$

$x.C$: Core distance w.r.t. $(\varepsilon, MinPts)$, $C_{\varepsilon, MinPts}(x)$

$x.R$: $\begin{cases} \min\limits_{p \in D} \{R_{\varepsilon, MinPts}(x, p)\}, & \text{if } x \text{ is non-core w.r.t. } (\varepsilon, MinPts) \\ \min\limits_{p \in D,\, p.P < x.P} \{R_{\varepsilon, MinPts}(x, p)\}, & \text{otherwise} \end{cases}$

$x.N$: Neighborhood size w.r.t. $\varepsilon$, $|N_\varepsilon(x)|$

$x.F$: $\begin{cases} x, & \text{if } x \text{ is noise w.r.t. } (\varepsilon, MinPts) \\ y \in N_\varepsilon(x) \text{ with max. } |N_\varepsilon(y)|, & \text{otherwise} \end{cases}$

Initially, we set $\min \emptyset = \infty$, and an object $o \in D$ is *non-core* w.r.t. $(\varepsilon, MinPts)$ if $|N_\varepsilon(o)| < MinPts$. Importantly, FINEX handles non-core objects w.r.t. $(\varepsilon, MinPts)$ differently, i.e., their reachability distance $x.R$ is globally minimized over *all* $p \in D$. This allows us to address the challenge of labeling *all* cluster objects correctly. Moreover, each object $x \in D$ features two additional attributes that are used to implement exact $MinPts^*$-queries (cf. Section 5.4): (1) $x.N$ denotes the neighborhood size, i.e., the number of neighbors. (2) $x.F$ is the *finder reference*, i.e., a reference to the (core) object that has the *most* neighbors and reaches $x$ w.r.t. $(\varepsilon, MinPts)$.

*Index Construction.* Algorithms 2 and 3 show the construction of our index. The overall structure of Algorithm 2 is reminiscent of the OPTICS algorithm [2]. However, updating the priority queue $Q$ differs substantially – cf. Algorithm 3.

In the outer loop of Algorithm 2, an arbitrary unprocessed object $o \in D$ is drawn and its core distance $o.C$ and neighborhood size $o.N$ are computed. Since $o$ has not been reached from any other object yet, we initialize $o.R$ to $\infty$ and the finder reference $o.F$ to refer to itself (implicitly). Afterwards, we append $o$ to the FINEX-ordering $\widetilde{O}$. If $o$ is a core object w.r.t. the generating pair $(\varepsilon, MinPts)$ (line 9), then all objects of $N_\varepsilon(o)$ are inserted into a priority queue $Q$ and the inner loop is executed (lines 11–17).

In each iteration, the inner loop pops the object $p \in Q$ with the smallest reachability distance (= highest priority), computes $p.C$ and $p.N$, and appends $p$ to $\widetilde{O}$. Note that $p.R$ and $p.F$ have already been computed while the priority queue $Q$ has been updated (cf. Algorithm 3). If $p$ is a core object too, all objects in $N_\varepsilon(p)$ are passed to Algorithm 3 (lines 16–17).

Algorithm 3 shows the update procedure of the priority queue $Q$. It iterates over all objects $q \in N_\varepsilon(c)$ for some core object $c$ (denoted $neighbors_c$). An object $o$ is marked as processed if $o \in \widetilde{O}$, whereas $o$ is unprocessed if $o \notin \widetilde{O}$. In contrast to OPTICS, we may remove non-core objects from the FINEX-ordering and reinsert them into $Q$ in order to globally minimize their reachability distance. Therefore, Algorithm 3 distinguishes three cases for $q$: (1) $q$ is unprocessed and $q \notin Q$ (line 3): The reachability attribute $q.R$ is set to $R_{\varepsilon,MinPts}(q, c)$ before $q$ is inserted into the queue. Note that $q.R$ represents the priority of $q$ (low reachability = high priority). (2) $q \in Q$ (and hence, $q$ is still unprocessed, line 7): If $R_{\varepsilon,MinPts}(q, c) < q.R$, then $q.R$ is updated and $q$'s priority is increased accordingly. (3) $q$ has already been processed (and hence, $q \notin Q$, line 11): We check (a) if $q$ is non-core w.r.t. $(\varepsilon, MinPts)$ and (b) if $R_{\varepsilon,MinPts}(q, c) < q.R$ (line 12). In this case, $q.R$ is updated accordingly and $q$ is removed from the FINEX-ordering $\widetilde{O}$, marked as unprocessed again, and reinserted into the queue $Q$. Effectively, case (3) globally minimizes the reachability distance $q.R$ of non-core objects (cf. Definition 5.1). Note that this does not change the asymptotic time complexity since every non-core object is reinserted into $Q$ at most $(MinPts - 1)$ times. Lines 16–17 are required to perform exact $MinPts^*$-queries and discussed in Section 5.4. Initialization of $o.F$ to a self-reference and $o.N$ to 0 for all $o \in D$ (cf. Algorithm 2) ensures that these lines are well-defined in all cases.

## 5.2 Linear-Time Clustering

One major drawback of OPTICS is the fact that border objects may be falsely labeled as noise. In this section, we describe how to mitigate this problem and highlight two advantages of FINEX over OPTICS: (i) Linear-time extraction of a *more accurate* clustering. (ii) For at least one parameter setting, i.e., $\varepsilon^* = \varepsilon$, FINEX can directly extract an *exact* clustering in linear time. Both are an immediate consequence of how we construct our index, i.e., the FINEX-ordering, and do not introduce overhead at runtime.

***More Accurate Clustering in Linear Time.*** After constructing FINEX, we can apply the cluster extraction technique of OPTICS (cf. Algorithm 1) to extract a more accurate density-based clustering (compared to the clustering extracted by OPTICS). Notably, replacing only the OPTICS-ordering with the FINEX-ordering already improves the approximate clustering substantially. For example, Figure 5 shows the resulting clusterings with the OPTICS- (Figure 5a) and the FINEX-ordering (Figure 5b), respectively. The global minimization of the reachability distance of non-core objects is reflected by smaller bars for $A$, $B$, $F$, and $G$ in Figure 5b. Alongside a new permutation on the x-axis of Figure 5b, FINEX correctly identifies *all* objects of the yellow cluster and three out of four objects of the blue cluster. This is a substantial improvement over the clustering extracted using the OPTICS-ordering, which only identifies 50% and 67% of the blue and the yellow cluster, respectively.

***Clustering of Border Objects.*** We now formalize the observations of Figure 5 by describing FINEX' approximate clusters and how FINEX improves the approximation provided by OPTICS.

---

**Algorithm 2:** FINEX-build($D$, $\varepsilon$, $MinPts$)

---

**Input:** dataset $D$, distance $\varepsilon$, count $MinPts$
**Output:** FINEX-ordering $\widetilde{O}$ w.r.t. ($\varepsilon$, $MinPts$)
`// o.F is initialized to self-reference` $\forall o \in D$
`// o.N is initialized to 0` $\forall o \in D$

1   $\widetilde{O} \leftarrow$ new empty ClusterOrdering
2   $Q \leftarrow$ new empty PriorityQueue
3   **while** $|\widetilde{O}| < |D|$ **do**
4      $o \leftarrow$ arbitrary unprocessed object in $D$
5      $o.C \leftarrow C_{\varepsilon, MinPts}(o)$
6      $o.N \leftarrow |N_\varepsilon(o)|$
7      $o.R \leftarrow \infty$
8      mark $o$ as processed and append it to $\widetilde{O}$
9      **if** $o.C \leq \varepsilon$ **then**
10        $Q.update(o, N_\varepsilon(o), \widetilde{O})$
11        **while** $Q \neq \varnothing$ **do**
12          $p \leftarrow Q.pop()$
13          $p.C \leftarrow C_{\varepsilon, MinPts}(p)$
14          $p.N \leftarrow |N_\varepsilon(p)|$
15          mark $p$ as processed and append it to $\widetilde{O}$
16          **if** $p.C \leq \varepsilon$ **then**
17            $Q.update(p, N_\varepsilon(p), \widetilde{O})$
18   **return** $\widetilde{O}$

---

**Algorithm 3:** PriorityQueue::update($c$, $neighbors_c$, $\widetilde{O}$)

---

**Input:** core object $c$, neighbors of $c$ $neighbors_c$, FINEX-ordering $\widetilde{O}$
`// note that` $c \in neighbors_c$

1   **foreach** $q \in neighbors_c$ **do**
2      $rdist \leftarrow R_{\varepsilon, MinPts}(q, c)$
3      **if** $q$ is unprocessed and not in PriorityQueue **then**
4        $q.R \leftarrow rdist$
5        $insert(q)$
6      **else**
7        **if** $q$ is in PriorityQueue **then**
8          **if** $rdist < q.R$ **then**
9            $q.R \leftarrow rdist$
10            $increasePriority(q)$
11        **else**
12          **if** $q.C > \varepsilon$ and $rdist < q.R$ **then**
13            mark $q$ as unprocessed and remove $q$ from $\widetilde{O}$
14            $q.R \leftarrow rdist$
15            $insert(q)$
16      **if** $c.N > q.F.N$ **then**
17        $q.F \leftarrow c$

---

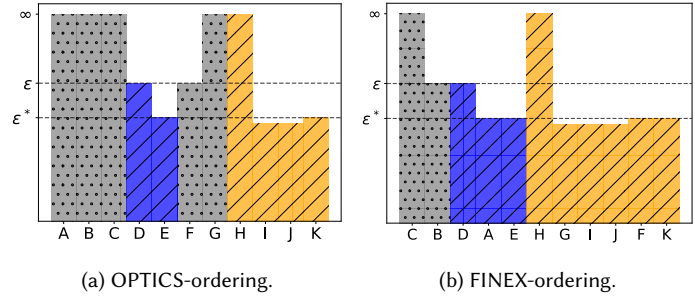

(a) OPTICS-ordering.　　　　(b) FINEX-ordering.

Fig. 5. Clusterings w.r.t. $\varepsilon^* = 3\varepsilon/4$ and $MinPts = 4$ for the dataset in Figure 4a as reported by Algorithm 1.

**Theorem 5.2 (Approximate FINEX Clusters).** *Let $\widetilde{O}_{\varepsilon,MinPts}$ be a FINEX-ordering computed by Algorithm 2. Considering an approximate cluster $S$ w.r.t. $(\varepsilon^*, MinPts)$ according to Definition 4.2, the following propositions hold:*

(a) $\exists$ *a cluster $K$ w.r.t. $(\varepsilon^*, MinPts)$ such that $S \subseteq K$.*

(b) $\forall p \in K : |N_\varepsilon(p)| \geq MinPts \implies p.P \leq x_n.P$.

(c) $\forall p \in K : |N_{\varepsilon^*}(p)| \geq MinPts \implies p \in S$.

Intuitively, $S$ is a subset of some density-based cluster $K$ w.r.t. $(\varepsilon^*, MinPts)$ that already contains all core objects of $K$. Due to (b), core objects w.r.t. $(\varepsilon, MinPts)$ that are border objects w.r.t. $(\varepsilon^*, MinPts)$ may be positioned before $x_1$ in the processing order. The fact that Algorithm 2 *globally* minimizes the reachability distance of non-core objects w.r.t. $(\varepsilon, MinPts)$ (as opposed to OPTICS) shows the strength of Theorem 5.2 and yields the following theorem:

**Theorem 5.3 (Clustering of Non-Cores).** *Let $x \in \widetilde{O}_{\varepsilon,MinPts}$ be a non-core object w.r.t. $(\varepsilon, MinPts)$. Given a distance threshold $\varepsilon^* \leq \varepsilon$, Algorithm 1 applied to $\widetilde{O}_{\varepsilon,MinPts}$ correctly assigns $x$ to*

- *the noise group if $x$ is a noise object w.r.t. $(\varepsilon^*, MinPts)$, or*
- *a corresponding cluster if $x$ is a border object w.r.t. $(\varepsilon^*, MinPts)$.*

Theorem 5.3 is based on the global minimization of the reachability distances and guarantees that border objects w.r.t. $(\varepsilon^*, MinPts)$ that are non-core objects w.r.t. $(\varepsilon, MinPts)$ *cannot* be falsely labeled as noise. This is not the case for OPTICS (cf. Section 4) and emphasizes the strength of our FINEX-ordering.

Another class of objects are so-called *former-cores*, i.e., core objects w.r.t. $(\varepsilon, MinPts)$ that become non-core w.r.t. $(\varepsilon^*, MinPts)$. Formally, an object $o \in D$ is a *former-core* w.r.t. $\varepsilon^* \leq \varepsilon$ and $MinPts$, if $|N_{\varepsilon^*}(o)| < MinPts$ and $|N_\varepsilon(o)| \geq MinPts$. Theorem 5.4 formalizes the fact that both OPTICS and FINEX correctly assign the same former-cores as border objects in their approximations.

**Theorem 5.4 (Clustering of Former-Cores).** *Let $x \in D$ be a former-core w.r.t. $\varepsilon^* \leq \varepsilon$ and $MinPts$, and let $\widetilde{O}_{\varepsilon,MinPts}$ and $O_{\varepsilon,MinPts}$ be the FINEX- and OPTICS-ordering, respectively. For $\varepsilon^*$, it holds that Algorithm 1 with $\widetilde{O}_{\varepsilon,MinPts}$ identifies $x$ as cluster member iff Algorithm 1 with $O_{\varepsilon,MinPts}$ identifies $x$ as cluster member.*

Note that Theorem 5.4 requires a stable priority queue w.r.t. insertion order, i.e., tied elements with equal priority are popped in insertion order. Consequently, reprocessing non-core objects w.r.t. $(\varepsilon, MinPts)$ in Algorithm 2 neither affects the reachability distance of other objects nor the overall order of former-cores (i.e., their order is identical in $\widetilde{O}_{\varepsilon,MinPts}$ and $O_{\varepsilon,MinPts}$). Combining Theorem 5.2, Theorem 5.3, and Theorem 5.4 yields that the approximate clustering of FINEX is *at least* as accurate as the approximate clustering of OPTICS.

*Exact Clustering in Linear Time.* The FINEX-ordering enables our solution to return an exact clustering w.r.t. $(\varepsilon^*, MinPts)$ in linear time for the special case $\varepsilon^* = \varepsilon$. This directly follows from Theorem 5.2 and Theorem 5.3, i.e., core and non-core objects are labeled correctly, and highlights another advantage of FINEX over OPTICS. We conclude with the corresponding corollary:

COROLLARY 5.5 (LINEAR EXACT QUERY). *Let $\widetilde{O}_{\varepsilon,MinPts}$ be a FINEX-ordering computed by Algorithm 2. Algorithm 1 with $\varepsilon^* = \varepsilon$ returns an exact clustering w.r.t. $(\varepsilon, MinPts)$ according to Definition 3.5 in linear time (in the number of objects).*

## 5.3 Exact $\varepsilon^*$-Queries

In Section 5.2, we discussed that FINEX returns an exact clustering for $\varepsilon^* = \varepsilon$. However, when $\varepsilon^* < \varepsilon$, some former-core objects may still be falsely labeled as noise (cf. object $C$ in Figure 5b). In this section, we introduce a new query type named $\varepsilon^*$-query that generalizes exact querying, such that FINEX is able to return exact clusterings for *any* $\varepsilon^* \leq \varepsilon$.

*$\varepsilon^*$-Query Algorithm.* Given our index and $\varepsilon^* \leq \varepsilon$, we proceed in two steps to efficiently derive an exact clustering w.r.t. $(\varepsilon^*, MinPts)$: (1) First, we obtain approximate clusters $S_1, \ldots, S_m$ w.r.t. $(\varepsilon^*, MinPts)$ using Algorithm 1. (2) For each $S_i$, we perform a *targeted search* for former-cores w.r.t. $\varepsilon^* \leq \varepsilon$ and $MinPts$ that are border objects of $S_i$. A former-core $o \in \widetilde{O}_{\varepsilon,MinPts}$ can be efficiently identified by checking $\varepsilon^* < o.C \leq \varepsilon$. A former-core $o$ is considered a *candidate* for $S_i$ when $o$ is labeled noise and is processed before the first object in $S_i$ (cf. Theorem 5.2). Candidates are subsequently verified by checking if the $\varepsilon^*$-neighborhood of $o$ contains a core object of $S_i$ w.r.t. $(\varepsilon^*, MinPts)$.

Proposition 3.9 further restricts the candidates, i.e., only former-cores of some sparse cluster $K$, $S_i \subseteq K$, become candidates because the core objects of $K$ surround the objects of $S_i$ in $\widetilde{O}_{\varepsilon,MinPts}$. The following theorem summarizes how to derive an exact clustering for $\varepsilon^* \leq \varepsilon$ and $MinPts$.

THEOREM 5.6 (EXACT $\varepsilon^*$-QUERY). *Let $\widetilde{O}_{\varepsilon,MinPts}$ be a FINEX-ordering computed by Algorithm 2. Algorithm 1 with $\varepsilon^* \leq \varepsilon$ returns approximate clusters $S_1, \ldots, S_m$ w.r.t. $(\varepsilon^*, MinPts)$. For each $S_i$, we add $o \in \widetilde{O}_{\varepsilon,MinPts}$ to $S_i$ if $o$ is directly density-reachable from some core object of $S_i$ w.r.t. $(\varepsilon^*, MinPts)$ and the following conditions hold:*

(1) *$o$ is labeled noise and $\varepsilon^* < o.C \leq \varepsilon$*
(2) *$o$ is processed before the first object in $S_i$*
(3) *$\exists$ a cluster $K$ w.r.t. $(\varepsilon, MinPts)$ such that $S_i \cup \{o\} \subseteq K$*
(4) *$o$ has not yet been added to another $S_j$, $j \neq i$*

*The result is an exact clustering w.r.t. $(\varepsilon^*, MinPts)$ according to Definition 3.5.*

*Discussion.* The conditions in Theorem 5.6 formally describe the constraints on the set of candidates. Conditions (1)–(3) can be derived as explained before. Condition (4) is required to assign ambiguous border objects to exactly one cluster.

Note that these tight restrictions on the candidates are only applicable to our FINEX-ordering and do *not* hold for OPTICS. Obviously, a FINEX-based $\varepsilon^*$-query requires fewer neighborhood computations than computations of DBSCAN from scratch. Furthermore, a single neighborhood computation for a candidate $o$ is more efficient due to two reasons: (i) In our context, a neighborhood computation only needs to consider the core objects of $S_i$ instead of all objects in $D$. (ii) We can terminate once a single core object is found in $o$'s $\varepsilon^*$-neighborhood, and do not need to compute exact distances to remaining cores. Overall, an $\varepsilon^*$-query is guaranteed to require fewer and (typically) more efficient neighborhood computations compared to DBSCAN from scratch.

### 5.4 Exact $MinPts^*$-Query

So far, we assumed $MinPts$ to be constant (similar to OPTICS). However, allowing to modify $MinPts$ increases the number of explorable clusterings. Therefore, we introduce the $MinPts^*$-*query*, which is a new query type that yields an exact clustering for constant $\varepsilon$ and *any* $MinPts^* \geq MinPts$. Recall that OPTICS has no support for $MinPts^*$-queries at all, hence we are left with DBSCAN as main competitor. FINEX' trick is, again, to leverage a fundamental property of density-based clusterings, i.e., dense clusters are contained in sparse clusters. Similar to Proposition 3.9 on $\varepsilon$-nested clusters, the following proposition formally defines $MinPts$-nested clusters.

PROPOSITION 5.7 ($MinPts$-NESTED CLUSTERS). *Let* $MinPts^* \geq MinPts$. *For every cluster* $K^* \subseteq D$ *w.r.t.* $(\varepsilon, MinPts^*)$ *there exists a cluster* $K \subseteq D$ *w.r.t.* $(\varepsilon, MinPts)$ *such that* $K^* \subseteq K$.

One key observation is that we can use the exact (sparse) clustering w.r.t. $(\varepsilon, MinPts)$ to filter out noise objects effectively, and hence, compute the clustering on a subset of the original dataset $D$. However, FINEX offers even more optimization options.

***$MinPts^*$-Query Algorithm.*** Given $MinPts^* \geq MinPts$, FINEX derives an exact clustering w.r.t. $(\varepsilon, MinPts^*)$ efficiently. Our $MinPts^*$-query consists of three steps: (1) Based on $\widetilde{O}_{\varepsilon, MinPts}$, we apply Algorithm 1 to obtain an exact clustering w.r.t. $(\varepsilon, MinPts)$ with $l$ sparse clusters $E_1, \ldots, E_l$ and discard all noise objects in $D$ that are not part of any $E_i$, $1 \leq i \leq l$. (2) For each $E_i$, $1 \leq i \leq l$, we identify density-connected sets of core objects w.r.t. $(\varepsilon, MinPts^*)$, i.e., we find all dense clusters within $E_i$ for $MinPts^*$. (3) Missing border objects w.r.t. $(\varepsilon, MinPts^*)$ are identified and added to the respective dense clusters using the corresponding finder references.

In step (1), we extract an exact clustering w.r.t. $(\varepsilon, MinPts)$ and do not consider $MinPts^* \geq MinPts$ yet. Consequently, we only filter out noise objects in $D$ w.r.t. $(\varepsilon, MinPts)$. In step (2), we identify clusters of higher density, i.e., $MinPts^* \geq MinPts$, within the sparse clusters $E_1, \ldots, E_l$. In other words, a sparse cluster may decompose into multiple dense clusters w.r.t. $MinPts^*$. Step (2) only involves core objects w.r.t. $\varepsilon$ and $MinPts^*$, identified via the neighborhood size attribute. The corresponding pseudocode is depicted in Algorithm 4 , where (partial) neighborhood computations take place on lines 6 and 12. Step (3) correctly assigns border objects to the dense clusters derived in step (2). This finally results in an exact clustering w.r.t. $(\varepsilon, MinPts^*)$.

***Discussion.*** Recall that the exact clustering assigns ambiguous border objects to exactly *one* cluster. Hence, for an exact cluster $E_i$ w.r.t. $(\varepsilon, MinPts)$, it holds that $E_i \subseteq K_i$ for some density-based cluster $K_i$ w.r.t. $(\varepsilon, MinPts)$. Consequently, applying standard DBSCAN with $(\varepsilon, MinPts^*)$ to the subset $E_i$ may yield incorrect results because core objects w.r.t. $(\varepsilon, MinPts^*)$ may lose their core status due to missing ambiguous border objects in $E_i$. We solve this by precomputing $o.N$ for all $o \in \widetilde{O}_{\varepsilon, MinPts}$ during index construction (cf. Algorithm 2). $o.N$ denotes the size of $N_\varepsilon(o) \subseteq D$ and therefore a core object $o \in E_i$ is correctly identified if $o.N \geq MinPts^*$. This also allows FINEX to completely avoid neighborhood computations if $o.N < MinPts^*$, which improves the efficiency of $MinPts^*$-queries.

Our $MinPts^*$-query also benefits from the fact that we only perform a neighborhood computation over the set of core objects w.r.t. $(\varepsilon, MinPts^*)$ to detect density-connected subsets of clusters w.r.t. $(\varepsilon, MinPts^*)$ in $E_i$. All border objects $o$ are subsequently added to these subsets based on the finder reference $o.F$ without involving any neighborhood computation. The finder reference $o.F$ is updated during index construction (cf. line 17 in Algorithm 3) whenever a core object with more neighbors is processed. Once Algorithm 2 terminates, $o.F$ references the densest neighbor of $o$.

---

**Algorithm 4:** ComputeCoreClustering(*Cores*)

---

**Input:** core objects w.r.t. ($\varepsilon, MinPts^*$) *Cores*

**Output:** collection of clustered core objects

1   $S \leftarrow \emptyset$                                                              `// S collects cluster objects`

2   $Stack \leftarrow \emptyset$

3   **while** $Cores \neq \emptyset$ **do**

4      $x \leftarrow Cores.pop()$

5      $S \leftarrow S \cup \{x\}$

6      $neighbors_x \leftarrow N_\varepsilon(x) \cap Cores$

7      $Cores \leftarrow Cores \setminus neighbors_x$

8      $Stack.push(neighbors_x)$

9      **while** $Stack \neq \emptyset$ **do**

10          $y \leftarrow Stack.pop()$

11          $S \leftarrow S \cup \{y\}$

12          $neighbors_y \leftarrow N_\varepsilon(y) \cap Cores$

13          $Cores \leftarrow Cores \setminus neighbors_y$

14          $Stack.push(neighbors_y)$

15      $Clustering \leftarrow Clustering \cup \{S\}$

16      $S \leftarrow \emptyset$

17   **return** $Clustering$

---

## 6   EMPIRICAL EVALUATION

In this section, we evaluate FINEX on 12 large real-world datasets from various domains and compare its clustering runtime to the exact baselines DBSCAN and AnyDBC [17] that compute clusterings from scratch. Moreover, we compare FINEX to the state-of-the-art indexing technique OPTICS in terms of build time and clustering accuracy when both indices are queried in linear runtime with Algorithm 1. Our empirical evaluation covers two different data types and distance functions: set data with the *Jaccard distance* and multi-dimensional vector data with the *Euclidean distance*.

***Datasets and Distance Functions.*** Our research is motivated by the process mining domain, where activities of event logs of business processes are modeled as sets of unique *tokens* [12, 15]. For instance, the sequence of events (***Start, Order, Pay, Return, reFund, End***) is modeled by the set $\{(S, O), (O, P), (P, R), (R, F), (F, E)\}$, where a tuple $(X, Y)$ in a set represents a transition from event $\boldsymbol{X}$ to $\boldsymbol{Y}$ [15]. For efficiency reasons, the transitions are bijectively mapped to integer tokens. We refer to this modeling approach as *set data*. For clustering sets, we use the Jaccard distance $d_J$, which is well-known from similarity search [3, 18, 25]. Given two sets $r, s \in D$, the Jaccard distance is defined as

$$d_J(r, s) = 1 - \frac{|r \cap s|}{|r \cup s|}$$

Intuitively, if $r$ and $s$ are similar, they share many tokens relative to their sizes, and thus $d_J$ will be small. Since set data in combination with the Jaccard distance is also of special interest to process mining companies, we use real-world datasets from this domain in our experiments: CELONIS-1/2 and CELONIS-NSP-1/2. In addition, we use four common, freely available datasets from the set similarity search domain, i.e., ENRON, REUTERS, FLICKR, and KOSARAK.

To demonstrate that FINEX generalizes to other data types and distance functions, we evaluate FINEX on four multi-dimensional vector datasets using the Euclidean distance, i.e., HOUSEHOLD, HT-SENSOR, PRECIPITATION, and GAS-SENSOR. The variables in our vector datasets have been standardized to zero mean and unit variance, a common procedure to ensure equal variable weights when working with the Euclidean distance [11, 22]. Table 2 lists all datasets and their characteristics.

Table 2. Characteristics of datasets in our experiments.

| | Dataset | Size | Deduplicated Size |
|---|---|---|---|
| **Sets** | CELONIS-1 | $4.9 \cdot 10^7$ | $8.2 \cdot 10^6$ |
| | CELONIS-2 | $1.6 \cdot 10^7$ | $2.6 \cdot 10^6$ |
| | CELONIS-NSP-1 | $1.4 \cdot 10^8$ | $1.2 \cdot 10^5$ |
| | CELONIS-NSP-2 | $1.2 \cdot 10^7$ | $6.5 \cdot 10^6$ |
| | ENRON [24] | $5.2 \cdot 10^5$ | $2.5 \cdot 10^5$ |
| | REUTERS [24] | $7.8 \cdot 10^5$ | $7.4 \cdot 10^5$ |
| | FLICKR [24] | $1.7 \cdot 10^6$ | $1.2 \cdot 10^6$ |
| | KOSARAK [24] | $9.9 \cdot 10^5$ | $6.1 \cdot 10^5$ |
| | **Dataset** | **Size** | **# Dimensions** |
| **Vectors** | PRECIPITATION [17] | $5.7 \cdot 10^5$ | 12 |
| | HOUSEHOLD [17] | $2.0 \cdot 10^6$ | 7 |
| | HT-SENSOR [20] | $9.3 \cdot 10^5$ | 10 |
| | GAS-SENSOR [17] | $4.2 \cdot 10^6$ | 16 |

**Data Deduplication**. As an optimization for set data, all compared algorithms leverage that datasets contain a large number of duplicated sets (cf. Table 2). Our implementations deduplicate sets while reading them into memory and store a duplicate count for each set. This ensures correct clustering results while the data is only partially materialized in memory. The duplicate counts are used to correctly determine the neighborhood sizes. Multi-dimensional vector data is unlikely to contain duplicates, hence we do not deduplicate our vector datasets.

**Implementation**. We implemented FINEX, DBSCAN, and OPTICS in a unified C++ framework. Moreover, the C++ code of AnyDBC [17] was thankfully provided by the original authors, and we adapted it to handle set data efficiently. Given a FINEX-ordering computed by Algorithm 2, an arbitrary number of clusterings can be queried with either $\varepsilon^* \leq \varepsilon$ (both exact and approximate) or $MinPts^* \geq MinPts$ (exact only). Analogously, a precomputed OPTICS-ordering can be queried with arbitrary $\varepsilon^* \leq \varepsilon$ to obtain approximate clusterings. The resulting clusterings are optionally saved to disk in the form of a table that maps each object to a cluster identifier while providing information about the core status of the respective object. Note that all algorithms are executed single-core.

**AnyDBC Configuration**. AnyDBC [17] requires configuring two additional input parameters $(\alpha, \beta)$ that define batch sizes for neighborhood computations. As proposed in the original publication [17], we globally set $\alpha = 512$ and $\beta = 4096$. Moreover, AnyDBC uses a random seed to select objects for which initial neighborhood computations are performed. To account for AnyDBC's random seed, all AnyDBC runtimes are means over five runs.

**Neighborhood Computations**. In terms of runtime, the most expensive part of density-based clustering is computing neighborhoods for the objects in $D$. For set data, we use a state-of-the-art *inverted list index* that leverages *prefix* and *length* filtering [21]. Our DBSCAN, OPTICS-build, and FINEX-build implementations perform the neighborhood computations as a separate step in advance, where all neighborhoods are materialized. Thereby, we leverage a *short prefix*, which is a set data-specific optimization that allows to partially compute the neighborhood for each $x \in D$. The full neighborhood of $x$ is derived by exploiting the symmetry of the Jaccard distance [18]. The neighborhood computations for FINEX' $MinPts^*$-query can be optimized in a similar manner. AnyDBC aims to perform neighborhood computations only for a subset of all objects, where each range query for an object $x \in D$ must return the full neighborhood of $x$. Hence, AnyDBC cannot leverage the short prefix optimization. Likewise, the short prefix is not applicable to FINEX' $\varepsilon^*$-query since candidate and core objects are disjoint.

Table 3. Average recall of border objects for FINEX and OPTICS over all datasets ($\varepsilon = 0.25$, $MinPts = 64$).

| $\varepsilon^*$ | 0.25 | 0.23 | 0.21 | 0.19 | 0.17 | 0.15 | 0.13 | 0.11 | 0.09 | 0.07 |
|---|---|---|---|---|---|---|---|---|---|---|
| FINEX | 1.000 | 0.977 | 0.966 | 0.957 | 0.949 | 0.936 | 0.932 | 0.923 | 0.909 | 0.884 |
| OPTICS | 0.944 | 0.942 | 0.945 | 0.944 | 0.943 | 0.935 | 0.932 | 0.923 | 0.909 | 0.884 |

Table 4. FINEX-build and OPTICS-build runtimes relative to DBSCAN from scratch ($\varepsilon = 0.25$, $MinPts = 64$).

| | FINEX-build | OPTICS-build |
|---|---|---|
| CELONIS-1 | 1.03 | 1.03 |
| CELONIS-2 | 1.01 | 1.02 |
| CELONIS-NSP-1 | 1.00 | 1.00 |
| CELONIS-NSP-2 | 1.00 | 1.00 |
| ENRON | 1.00 | 1.00 |
| REUTERS | 0.97 | 0.97 |
| FLICKR | 1.12 | 1.10 |
| KOSARAK | 1.03 | 1.02 |
| PRECIPITATION | 1.23 | 1.17 |
| HOUSEHOLD | 1.60 | 1.39 |
| HT-SENSOR | 1.25 | 1.16 |
| GAS-SENSOR | 1.16 | 1.12 |

In all implementations for vector data, neighborhood computations are facilitated by the in-memory, cache-friendly *kd-tree* proposed by Kennel [14]. Neighborhoods are not materialized in advance for vector data. Note that all reported clustering runtimes include the runtime of the respective neighborhood computations.

***Machine Setup***. We use a 64-bit machine equipped with 2 physical Intel Xeon E5-2603 v4 CPUs, 1.70GHz, 6 cores each. The cores share a 15 MiB L3 cache and have another 256 KiB of independent L2 cache. The system has 96 GiB of RAM and runs Debian Buster (Linux 4.19.0-14-amd64 #1 SMP Debian 4.19.171-2 (2021-01-30)).

## 6.1 Index Comparison

In this section, we compare FINEX to the state-of-the-art OPTICS index in terms of $\varepsilon^*$-query accuracy and index build time.

***Accuracy***. Both OPTICS and FINEX can be queried in linear time using Algorithm 1 to retrieve an approximate clustering. From Section 5.2, we know that FINEX approximates the exact clustering better than OPTICS does. To illustrate this, Table 3 summarizes the average recall of border objects for FINEX and OPTICS over all datasets and varying $\varepsilon^* \leq \varepsilon$ ($\varepsilon = 0.25$, $MinPts = 64$). Compared to FINEX, OPTICS misses up to 5.6% of border objects on average, and the difference is more significant for higher $\varepsilon^*$ values. Due to Theorem 5.3, FINEX correctly assigns all border objects w.r.t. $\varepsilon^*$ that are also border objects w.r.t. $\varepsilon$. This guarantee does not hold for OPTICS. Note that FINEX is 100% accurate for $\varepsilon^* = \varepsilon$. With decreasing $\varepsilon^*$, the recalls of FINEX and OPTICS converge, as the number of non-core border objects declines until the only border objects left are former-cores, which are treated equally by both indices (cf. Theorem 5.4).

***Build Time.*** Similar to DBSCAN, the runtime complexities of FINEX-build (Algorithm 2) and OPTICS-build are dominated by the complexity of neighborhood computations. Therefore, all three algorithms have the same asymptotic time complexity. Table 4 depicts runtimes of FINEX-build and OPTICS-build relative to DBSCAN from scratch for $\varepsilon = 0.25$ and $MinPts = 64$. Values smaller (larger) than 1 signify that building the respective index is faster (slower) than executing DBSCAN from scratch. We observe that FINEX-build, OPTICS-build, and DBSCAN perform almost identical on set data. On vector data, however, FINEX-build and OPTICS-build are up to 60% and 39% slower than DBSCAN, respectively. We attribute this to the priority queue operations of FINEX-build and OPTICS-build in combination with the higher share of core objects that modify the queue (the average share of core objects is 85.8 % on our vector datasets and 46.2 % on our set datasets, $\varepsilon = 0.25$ and $MinPts = 64$, respectively). Furthermore, FINEX performs more insert and pop operations on the queue, which results in slightly higher build times compared to OPTICS, i.e., between 4% and 21% (w.r.t. DBSCAN from scratch).

## 6.2 $\varepsilon^*$-Queries

In this section, we evaluate $\varepsilon^*$-queries on set and vector data. The generating $(\varepsilon, MinPts)$ pair is fixed with $\varepsilon = 0.25$ and $MinPts = 64$ for both Jaccard and Euclidean distance. We query $\varepsilon^* \leq \varepsilon$ with

$$\varepsilon^* \in \{0.25,\ 0.23,\ 0.21,\ 0.19,\ 0.17,\ 0.15,\ 0.13,\ 0.11,\ 0.09,\ 0.07\}.$$

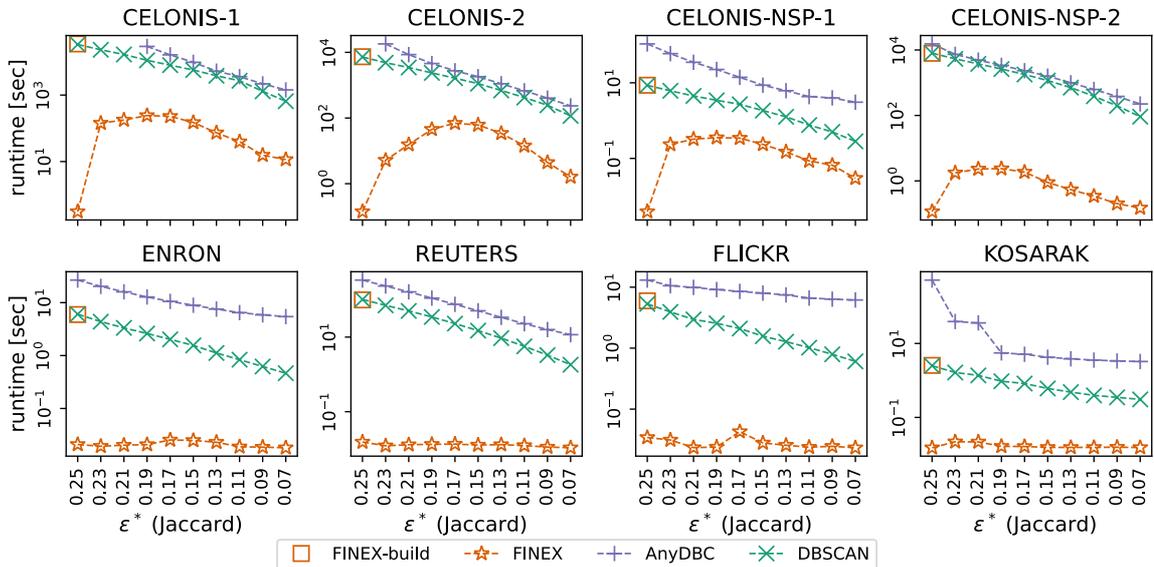

Fig. 6. Clustering runtime over $\varepsilon^* \leq \varepsilon$ for set data (Jaccard distance; generating $\varepsilon = 0.25$ and $MinPts = 64$).

***Results on Set Data.*** In Figure 6, we compare the runtime of exact FINEX clustering to its competitors DBSCAN and AnyDBC. FINEX outperforms its competitors in all settings, often by multiple orders of magnitude. The runtimes of DBSCAN and AnyDBC decrease for lower values of $\varepsilon^*$ because the prefix-based neighborhood computations become more efficient for lower distance thresholds. AnyDBC performs worst on set data, for which we identify two reasons: (i) AnyDBC fails to effectively prune neighborhood computations when combined with the Jaccard distance. The median of avoided neighborhood computations is 0.4% over all AnyDBC runs shown in Figure 6 (min = 0%, max = 18.4%). We attribute this to the upper bound of $3\varepsilon^*$ that AnyDBC uses to separate core objects, which is less effective for the Jaccard distance. (ii) AnyDBC cannot use the short prefix optimization described in the beginning of Section 6, which is available for DBSCAN. The plots for CELONIS-1 and CELONIS-2 miss data points for AnyDBC and high $\varepsilon^*$ values since the execution terminated after a timeout of 12 hours.

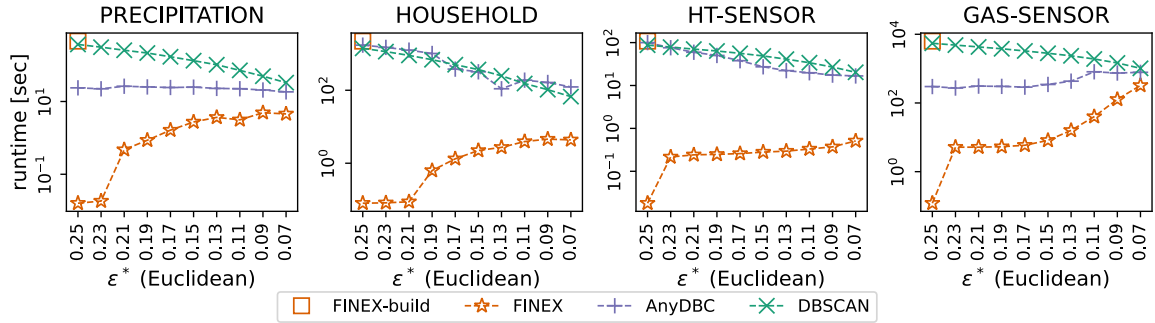

Fig. 7. Clustering runtime over $\varepsilon^* \leq \varepsilon$ for vector data (Euclidean distance; generating $\varepsilon = 0.25$ and $MinPts = 64$).

FINEX is most efficient when $\varepsilon^* = \varepsilon = 0.25$. In this case, only a linear scan through the FINEX-ordering is necessary to obtain an exact clustering. In all other cases, an additional candidate-verification step (cf. Section 5.3) is required to obtain an exact clustering. Moreover, Figure 6 shows that FINEX' runtime is bell-shaped. This is caused by the matching of candidates to core objects, which involves $O(n_k \cdot n_c)$ verifications, where $n_k$ is the number of candidates and $n_c$ is the number of cores. However, according to Theorem 5.6, each candidate is a former core object. Hence, when $n_k$ is low, then $n_c$ is high (and vice versa). As a consequence, querying is most expensive when both $n_k$ and $n_c$ are of medium size.

***Results on Vector Data.*** In Figure 7, AnyDBC performs better than DBSCAN on the datasets PRECIPITATION and GAS-SENSOR, while they behave similarly for HOUSEHOLD and HT-SENSOR. In contrast to set data, AnyDBC achieves to effectively prune neighborhood computations. The median of avoided neighborhood computations is 48.6% over all AnyDBC runs shown in Figure 7 (min = 17%, max = 90.9%), which explains why AnyDBC is the superior baseline algorithm for vector data. FINEX, however, outperforms both competitors on all vector datasets and parameter settings with runtime difference of up to three orders of magnitude (cf. HOUSEHOLD). The least difference between the algorithms can be observed for the GAS-SENSOR dataset for smaller values of $\varepsilon^*$. This is explained by the fact that GAS-SENSOR almost exclusively consists of core objects (more than 99.8%) at the generating distance of $\varepsilon = 0.25$. Thus, many candidates are formed when $\varepsilon^*$ is reduced and FINEX executes a relatively large number of neighborhood computations to verify these candidates against remaining cores.

### 6.3 *MinPts*\*-Queries

Next, we evaluate *MinPts*\*-queries on set and vector data. The generating ($\varepsilon$, *MinPts*) pair is fixed with $\varepsilon = 0.15$ and *MinPts* = 16 for both Jaccard and Euclidean distance. We query *MinPts*\* $\geq$ *MinPts* with

$$MinPts^* \in \{16, \ 32, \ 64, \ 128, \ 256, \ 512, \ 1024\}.$$

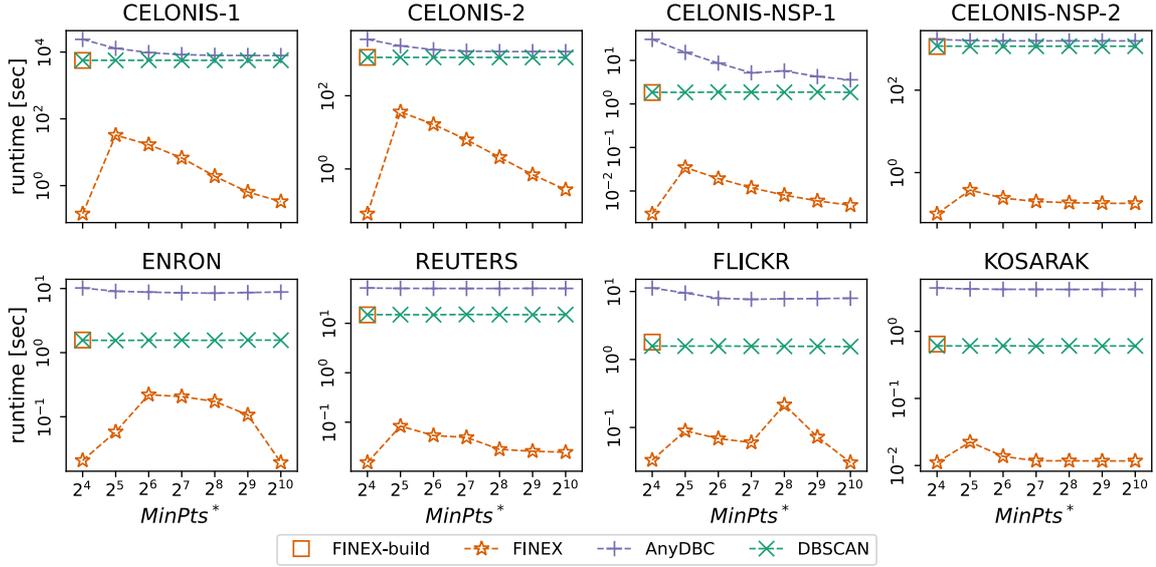

**Fig. 8.** Clustering runtime over *MinPts*\* $\geq$ *MinPts* for set data (Jaccard distance; generating $\varepsilon = 0.15$ and *MinPts* = 16).

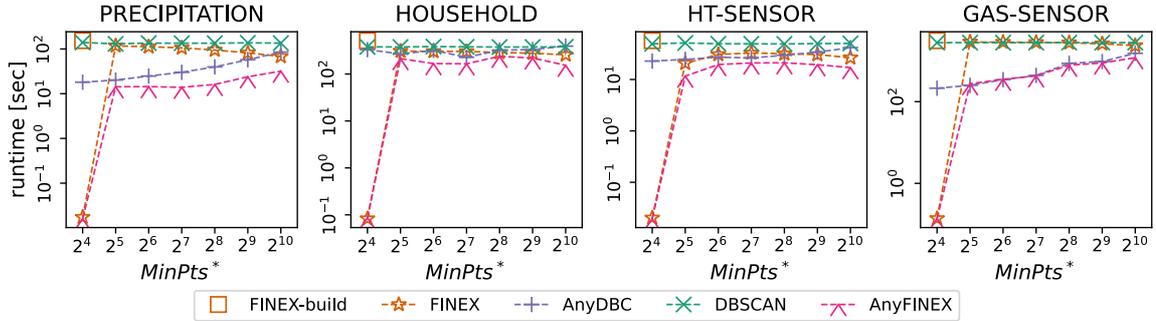

**Fig. 9.** Clustering runtime over *MinPts*\* $\geq$ *MinPts* for vector data (Euclidean distance; generating $\varepsilon = 0.15$ and *MinPts* = 16).

***Results on Set Data***. Figure 8 shows that FINEX outperforms DBSCAN and AnyDBC in all investigated scenarios. Even in the worst case (ENRON and REUTERS), we observe runtime improvements of approximately one order of magnitude over DBSCAN. Note that DBSCAN's runtime is insensitive to changes of *MinPts*\* since the computational complexity of neighborhood computations only depends on $\varepsilon$. This does not necessarily apply to AnyDBC, which takes more time than FINEX and DBSCAN on set data. Again, AnyDBC suffers from the same limitations outlined in Section 6.2. The median of avoided neighborhood computations is 0.2% over all AnyDBC runs on set data shown in Figure 8, whereas it is 43.8% over all AnyDBC runs on vector data shown in Figure 9. FINEX is most efficient when $MinPts^* = MinPts = 16$ since a linear scan over the FINEX-ordering yields an exact clustering (cf. Corollary 5.5).

For $MinPts^* \geq MinPts$, the computational complexity of FINEX' $MinPts^*$-queries is determined by the search for density-connected core objects (Algorithm 4). The number of core objects decreases when $MinPts^*$ increases. Hence, the computational complexity is also expected to decrease for increasing $MinPts^*$. We observe exceptions from this rule for ENRON at $MinPts^* = 2^6$ and FLICKR at $MinPts^* = 2^8$, where the runtime increases. We attribute this to an optimization of our $MinPts^*$-query: density-connected cores only have to be determined when an object $x \in D$ with $MinPts^* > |N_\varepsilon(x)| \geq MinPts$ exists. If no such object exists, all cores w.r.t. $(\varepsilon, MinPts)$ preserve their core status w.r.t. $(\varepsilon, MinPts^*)$. Thus, density-connected components can be directly obtained from the sparse clustering in the FINEX-ordering. In the mentioned cases, the number of preserved cores decreases substantially, which results in additional expensive searches for density-connected components.

**Results on Vector Data.** Figure 9 shows our runtime results for $MinPts^*$-queries on vector data. We observe that AnyDBC outperforms FINEX on GAS-SENSOR and PRECIPITATION (for low $MinPts$ values). For HOUSEHOLD and HT-SENSOR, AnyDBC and FINEX have comparable runtimes. Furthermore, AnyDBC outperforms DBSCAN in almost all scenarios, making AnyDBC the favorable baseline for $MinPts^*$-queries. Therefore, we additionally evaluate *AnyFINEX*, a proof-of-concept implementation of a FINEX variation that uses AnyDBC to identify density-connected components (instead of the DBSCAN-like Algorithm 4). In a nutshell, AnyFINEX combines FINEX' noise filter with a faster baseline to reduce the overall number of neighborhood computations. On PERCIPITATION, for instance, AnyFINEX' benefits are twofold: (a) Low runtime for small $MinPts$ values since AnyDBC is used to identify density-connected components. (b) The decreasing runtime of FINEX for high $MinPts$ values due to noise filtering. In contrast, the runtime of AnyDBC increases with $MinPts$ because it does not leverage the noise filtering of FINEX. On GAS-SENSOR, FINEX is hardly able to filter out noise objects, resulting in comparable runtimes for AnyFINEX (resp. FINEX) and AnyDBC (resp. DBSCAN). Overall, our results empirically confirm that AnyFINEX integrates the benefits of AnyDBC *and* FINEX. Since DBSCAN is faster than AnyDBC on set data, we do not expect AnyFINEX to provide runtime benefits on sets.

## 7 CONCLUSIONS

We studied the problem of identifying exact density-based clusterings for changing parameters $\varepsilon$ and $MinPts$. We proposed FINEX  the first linear-space index that overcomes the shortcomings of existing solutions, i.e., efficiency, flexibility, and support for changing the input parameters. FINEX is constructed for a generating pair $(\varepsilon, MinPts)$ and is able to return an *exact* density-based clustering w.r.t. either $\varepsilon^* \leq \varepsilon$ or $MinPts^* \geq MinPts$. In addition, the FINEX-ordering enables our index to generate more accurate clusterings (compared to OPTICS) as well as the exact clustering for $\varepsilon^* = \varepsilon$, both in linear time. Furthermore, FINEX exploits basic properties of density-based clusters to effectively prune expensive neighborhood computations. In our experiments on 12 large real-world datasets with set-valued and multi-dimensional vector objects, FINEX outperforms its competitors often by orders of magnitude.

**Future Work.** Unlike OPTICS [1], FINEX is currently not able to incorporate updates to the underlying dataset, i.e., when objects are added to or removed from the dataset the index must be rebuilt. Hence, we plan to investigate efficient update strategies for FINEX.

FINEX is designed as an efficient in-memory index. The linear (sequential) scan over FINEX to retrieve core and candidate objects for $\varepsilon^*$-queries is also favorable for an external-memory extension. After storing these objects consecutively on external memory, they can be scanned for the neighborhood computations. Efficient retrieval of neighborhoods is an orthogonal problem, and FINEX will benefit from existing external-memory neighborhood indexes. *MinPts*-queries are implemented in a similar manner. Extending and testing FINEX as an external-memory index poses an interesting direction for future research.

## ACKNOWLEDGMENTS

This research was funded in whole, or in part, by the Austrian Science Fund (FWF) P 34962. For the purpose of open access, the authors have applied a CC BY public copyright license to any author accepted manuscript version arising from this submission.

## REFERENCES

[1] Elke Achtert, Christian Böhm, Hans-Peter Kriegel, and Peer Kröger. 2005. Online Hierarchical Clustering in a Data Warehouse Environment. In *Proceedings of the Fifth IEEE International Conference on Data Mining (ICDM '05)*. IEEE Computer Society, USA.

[2] Mihael Ankerst, Markus M. Breunig, Hans-Peter Kriegel, and Jörg Sander. 1999. OPTICS: Ordering Points To Identify the Clustering Structure. In *Proceedings of ACM SIGMOD Int. Conf. on Management of Data (SIGMOD '99, Vol. 28)*. ACM New York, NY, USA, 49–60.

[3] Nikolaus Augsten and Michael Bohlen. 2013. *Similarity Joins in Relational Database Systems* (3rd ed.). San Rafael: Morgan & Claypool Publishers.

[4] Jon Louis Bentley. 1975. Multidimensional binary search trees used for associative searching. *Commun. ACM* 18, 9 (1975), 509–517.

[5] Stefan Brecheisen, Hans-Peter Kriegel, and Martin Pfeifle. 2006. Parallel Density-Based Clustering of Complex Objects. In *Advances in Knowledge Discovery and Data Mining (PAKDD '06)*. Springer Berlin Heidelberg, 179–188.

[6] Ricardo J.G.B. Campello, Davoud Moulavi, and Jörg Sander. 2013. Density-Based Clustering Based on Hierarchical Density Estimate. In *Advances in Knowledge Discovery and Data Mining (PAKDD '13)*. Springer Berlin Heidelberg, 160–172.

[7] Paolo Ciaccia, Marco Patella, and Pavel Zezula. 1997. M-tree: an Efficient Access Method for Similarity Search in Metric Spaces. In *Proceedings of the 23rd International Conference on Very Large Data Bases (VLDB '97)*. Morgan Kaufmann Publishers Inc., San Francisco, CA, USA, 426–435.

[8] Martin Ester, Hans-Peter Kriegel, Jörg Sander, and Xiaowei Xu. 1996. A Density-Based Algorithm for Discovering Clusters in Large Spatial Databases with Noise. In *Proceedings of the Second International Conference on Knowledge Discovery and Data Mining (KDD '96)*. AAAI Press, 226–231.

[9] Junhao Gan and Yufei Tao. 2018. Fast Euclidean OPTICS with Bounded Precision in Low Dimensional Space. In *Proceedings of the 2018 International Conference on Management of Data (SIGMOD '18)*. Association for Computing Machinery, New York, NY, USA, 1067–1082.

[10] Antonin Guttmann. 1984. R-Trees: A Dynamic Index Structure For Spatial Searching. In *Proceedings of the 1984 ACM SIGMOD International Conference on Management of Data (SIGMOD '84, Vol. 14)*. Association for Computing Machinery, New York, NY, USA, 47–57.

[11] Jiawei Han, Micheline Kamber, and Jian Pei. 2012. *Data Mining: Concepts and Techniques* (3rd edition ed.). Morgan Kaufmann.

[12] B. F. A. Hompes, J. C. A. M. Buijs, W. M. P. van der Aalst, P. M. Dixit, and J Buurman. 2015. Discovering Deviating Cases and Process Variants Using Trace Clustering. In *27th Benelux Conference on Artificial Intelligence (BNAIC 2015)*. Hasselt, Belgium.

[13] Jennifer Jang and Heinrich Jiang. 2019. DBSCAN++: Towards fast and scalable density clustering. In *Proceedings of the 36th International Conference on Machine Learning*, Vol. 97. PMLR, 3019–3029.

[14] Matthew B Kennel. 2004. KDTREE 2: Fortran 95 and C++ software to efficiently search for near neighbors in a multi-dimensional Euclidean space. *arXiv preprint physics/0408067* (2004).

[15] Daniel Kocher, Nikolaus Augsten, and Willi Mann. 2021. Scaling Density-Based Clustering to Large Collections of Sets. In *Proceedings of the 24th International Conference on Extending Database Technology*. EDBT, 109–120.

[16] Son T Mai, Ira Assent, and Martin Storgaard. 2016. AnyDBC: An efficient anytime density-based clustering algorithm for very large complex datasets. In *Proceedings of the 22nd ACM SIGKDD international conference on knowledge discovery and data mining*. 1025–1034.

[17] Son T. Mai, Jon Jacobsen, Sihem Amer-Yahia, Ivor Spence, Nhat-Phuong Tran, Ira Assent, and Quoc Viet Hung Nguyen. 2022. Incremental Density-Based Clustering on Multicore Processors. *IEEE Transactions on Pattern Analysis and Machine Intelligence* 44, 3 (2022), 1338–1356.

[18] Willi Mann, Nikolaus Augsten, and Panagiotis Bouros. 2016. An Empirical Evaluation of Set Similarity Join Techniques. In *Proceedings of the VLDB Endowment*, Vol. 9. San Francisco, CA, USA, 636–647.

[19] Natalie Price-Jones and Jo Bovy. 2019. Blind chemical tagging with DBSCAN: prospects for spectroscopic surveys. *Monthly Notices of the Royal Astronomical Society* 487, 1 (2019), 871–886.

[20] UCI Machine Learning Repository. 2022. HT-SENSOR. http://archive.ics.uci.edu/ml/datasets/gas+sensors+for+home+activity+monitoring. Accessed: October 2022.

[21] Leonardo Andrade Ribeiro and Theo Härder. 2009. Efficient Set Similarity Joins Using Min-prefixes. In *Proceedings of the 13th East-European Conference on Advances in Databases and Information Systems*. ADBIS, 88–102.

[22] scikit learn. 2022. Standardization, or mean removal and variance scaling. https://scikit-learn.org/stable/modules/preprocessing.html. Accessed: October 2022.

[23] Kevin Sheridan, Tejas G Puranik, Eugene Mangortey, Olivia J Pinon-Fischer, Michelle Kirby, and Dimitri N Mavris. 2020. An application of dbscan clustering for flight anomaly detection during the approach phase. In *AIAA Scitech 2020 Forum*.

[24] Xubo Wang, Lu Qin, Xuemin Lin, Ying Zhang, and Lijun Chang. 2019. Leveraging set relations in exact and dynamic set similarity join. In *The VLDB Journal*. 267–292.

[25] Chuan Xiao, Wei Wang, Xuemin Lin, Jeffrey Xu Yu, and Guoren Wang. 2011. Efficient Similarity Joins for Near-Duplicate Detection. *ACM Transactions on Database Systems* 36, 3 (2011).

## A  PROOFS

In this section, we provide formal proofs for all theoretical results of this paper.

PROPOSITION 3.9 ($\varepsilon$-NESTED CLUSTERS). *Let* $0 \leq \varepsilon^* \leq \varepsilon$. *For every cluster* $K^* \subseteq D$ *w.r.t.* ($\varepsilon^*$, *MinPts*) *there is a cluster* $K \subseteq D$ *w.r.t.* ($\varepsilon$, *MinPts*) *such that* $K^* \subseteq K$.

PROOF. Let $K^* \subseteq D$ be a cluster w.r.t. $\varepsilon^*$ and *MinPts* and let $p \in K^*$ be a core object w.r.t. ($\varepsilon^*$, *MinPts*). Since $\varepsilon^* \leq \varepsilon$ it follows that

$$N_{\varepsilon^*}(p) = \{x \in D : d(x, p) \leq \varepsilon^*\} \subseteq \{x \in D : d(x, p) \leq \varepsilon\} = N_\varepsilon(p)$$

and consequently

$$MinPts \leq |N_{\varepsilon^*}(p)| \leq |N_\varepsilon(p)|$$

By Lemma 1 in [8], the set

$$K \coloneqq \{x \in D : x \text{ is density-reachable from } p \text{ w.r.t. } \varepsilon \text{ and } MinPts\}$$

is a cluster in $D$ w.r.t. $\varepsilon$ and *MinPts*. Now, let $q \in K^*$ arbitrary. From connectivity in Definition 3.4 it follows that $q$ is density-reachable from $p$ w.r.t. $\varepsilon^*$ and *MinPts* (remember that $p$ is a core object of $K^*$). That is, there exist $p_1, \ldots, p_n \in D$, $p_1 = p$, $p_n = q$ with

$$p_{i+1} \in N_{\varepsilon^*}(p_i) \ \wedge \ |N_{\varepsilon^*}(p_i)| \geq MinPts \ \forall i \in \{1, \ldots, n-1\}$$

Though, it holds that $N_{\varepsilon^*}(p_i) \subseteq N_\varepsilon(p_i)$ for all $i$. Thus

$$p_{i+1} \in N_\varepsilon(p_i) \ \wedge \ |N_\varepsilon(p_i)| \geq MinPts \ \forall i \in \{1, \ldots, n-1\}$$

Consequently, $q$ is density-reachable from $p$ w.r.t. $\varepsilon$ and *MinPts*. It follows that $q \in K$. ☐

---

The following two lemmas provide additional properties of cluster orderings. They will be deployed in the proof for Theorem 4.3 (in the form of proof by contradiction).

LEMMA A.1 (EFFECTS OF DIRECT DENSITY-REACHABILITY). *Let* $O_{\varepsilon, MinPts}$ *be an OPTICS-ordering and* $\varepsilon^* \in \mathbb{R}_{\geq 0}$ *with* $\varepsilon^* \leq \varepsilon$. *Moreover, let* $x, y \in O_{\varepsilon, MinPts}$ *such that:*

(1) $y$ *is directly density-reachable from* $x$ *w.r.t.* $\varepsilon^*$
(2) $x.P < y.P$

*Then, it holds* $\forall o \in O_{\varepsilon, MinPts}$ *with* $x.P < o.P \leq y.P$ *that* $o.R \leq \varepsilon^*$.

PROOF. From $x.P < y.P$, we know that at the time when $x$ was processed, $y$ has not been processed yet. Since $x$ is a core object at $\varepsilon^*$ and $d(x, y) \leq \varepsilon^*$, it holds that $y \in Q$ with $y.R \leq \varepsilon^*$ after $x$ was processed. From this time on, the inner loop was executed (at least) until $y$ has been processed (priority queue $Q$ nonempty). While $y$ was in $Q$, $y.R$ could have been decreased further, but it could not have been increased. That is, when $y$ was processed, it was still true that $y.R \leq \varepsilon^*$. Moreover, the inner OPTICS loop selects the element $o$ with minimum $o.R$ among all elements in $Q$. It follows for all objects $o$ that were processed between $x$ and $y$ that $o.R \leq y.R \leq \varepsilon^*$. □

Intuitively, Lemma A.1 states that if an object $y$ is directly density-reachable from another object $x$ w.r.t. some $\varepsilon^* \leq \varepsilon$, and $x$ was processed before $y$, then all objects processed in-between are assigned a reachability distance of at most $\varepsilon^*$. An analogous property holds for the more general concept of density-reachability.

LEMMA A.2 (EFFECTS OF DENSITY-REACHABILITY). *Let* $O_{\varepsilon, MinPts}$ *be an OPTICS-ordering and* $\varepsilon^* \in \mathbb{R}_{\geq 0}$ *with* $\varepsilon^* \leq \varepsilon$. *Moreover, let* $x_1, \dots, x_n \in O_{\varepsilon, MinPts}$ *such that:*

(1) $x_n$ *is density-reachable from* $x_1$ *w.r.t.* $\varepsilon^*$ *via* $x_2, \dots, x_{n-1}$
(2) *for all* $i = 2, \dots, n$, *it holds that* $x_1.P < x_i.P$

*Then, it holds* $\forall o \in O_{\varepsilon, MinPts}$ *with* $x_1.P < o.P \leq x_n.P$ *that* $o.R \leq \varepsilon^*$.

PROOF. We show the proposition by induction.

$\underline{n = 2}$: Directly follows from Lemma A.1.

$\underline{n \rightarrow n + 1}$: We distinguish two cases.

(i) $x_n.P < x_{n+1}.P$

By assumption $x_{n+1}$ must be directly density-reachable from $x_n$ w.r.t. $\varepsilon^*$. Thus, Lemma A.1 yields

$$\forall o \in O_{\varepsilon, MinPts} \text{ with } x_n.P < o.P \leq x_{n+1}.P \; : \; o.R \leq \varepsilon^*$$

Moreover, from the induction hypothesis, we know

$$\forall o \in O_{\varepsilon, MinPts} \text{ with } x_1.P < o.P \leq x_n.P \; : \; o.R \leq \varepsilon^*$$

Combining both yields the proposition.

(ii) $x_n.P > x_{n+1}.P$

The proposition directly follows from the induction hypothesis.

□

In other words, if $x_n$ is density-reachable from $x_1$ w.r.t. $\varepsilon^*$, and $x_1$ is processed first among the chosen objects that connect it to $x_n$, then all objects processed in-between are assigned a reachability distance of at most $\varepsilon^*$. Note that there is no processing order imposed on the chain of objects other than $x_1$ being the first one. E.g., it is not required that $x_n$ is the last object processed for the Lemma A.2 to hold.

---

THEOREM 4.3 (APPROXIMATE OPTICS CLUSTERS). *Let* $O_{\varepsilon, MinPts}$ *be a cluster ordering computed by OPTICS. We consider an approximate cluster $S$ w.r.t.* $(\varepsilon^*, MinPts)$ *according to Definition 4.2. The following propositions hold:*

(a) $\exists$ *a density-based cluster $K$ w.r.t.* $\varepsilon^*$ *and $MinPts$ according to Definition 3.4 such that $S \subseteq K$.*
(b) $\forall p \in K : p.P \leq x_n.P$
(c) $\forall p \in K : |N_{\varepsilon^*}(p)| \geq MinPts \implies p \in S$

Proof. **Proposition (a):** By assumption, $x_1 \in S$ is a core object at $\varepsilon^*$ and by Lemma 1 in [8], the set

$$K := \{p \in O_{\varepsilon, MinPts} : p \text{ is density-reachable from } x_1 \text{ w.r.t. } \varepsilon^* \text{ and } MinPts\}$$

is a cluster. We proceed by showing the following proposition by strong induction:

$$\forall i \in \{1, \ldots, n\} : x_i \in S \text{ is density-reachable from } x_1 \in S \text{ w.r.t. } \varepsilon^* \text{ and } MinPts$$

$\underline{i = 1}$: Trivially holds.

$\underline{i = 1, \ldots, k \rightarrow i = k + 1}$: By assumption, we have $x_1.R > \varepsilon^*$, i.e., when $x_1$ was processed it must have held in the priority queue $Q$ that

$$\forall p \in Q : \varepsilon^* < x_1.R \leq p.R$$

Otherwise, $x_1$ would not have been picked. Note that this includes the situation when $Q$ was empty (then, $x_1$ would have been processed in the outer OPTICS loop and $x_1.R = \infty$). Consequently, $x_{k+1}$ could only have been processed with $x_{k+1}.R \leq \varepsilon^*$ if it was inserted in $Q$ and/or decreased by some of its predecessors in $S$. Let $x_m$, $1 \leq m \leq k$ be the last object that inserted or decreased $x_{k+1}$. Then, the reachability distance

$$R_{\varepsilon, MinPts}(x_{k+1}, x_m) = x_{k+1}.R \leq \varepsilon^*$$

That is, $x_{k+1}$ is directly density-reachable from $x_m$ w.r.t. $\varepsilon^*$ and $MinPts$. Combining this with the induction hypothesis yields the desired proposition.

We've shown that for all $x_i \in S$ it holds that $x_i \in K$. □

**Propositions (b) and (c):** Assume there exists some core objects $p$ of $K$ processed before $x_1 \in S$. Then, there must exist a leftmost core object $c \in K$. Thus, $x_1$ is density-reachable from $c$ w.r.t. $\varepsilon^*$ and $MinPts$ via some objects processed after $c$. Lemma A.2 yields

$$\forall o \in O_{\varepsilon, MinPts} \text{ with } c.P < o.P \leq x_1.P : o.R \leq \varepsilon^*$$

In particular, $x_1.R \leq \varepsilon^*$, which contradicts property (2) in Definition 4.2. It follows that there are no core objects of $K$ processed before $x_1$.

Let's assume now that there exists an object $p \in K$ with $p.P > x_n.P$ (i.e., an object processed after $x_n \in S$). $p$ must be density-reachable from $x_1$ w.r.t. $\varepsilon^*$ and $MinPts$ (according to Definition 3.4). Since $x_1$ is the leftmost core object, we can again apply Lemma A.2. It yields

$$\forall o \in O_{\varepsilon, MinPts} \text{ with } x_1.P < o.P \leq p.P : o.R \leq \varepsilon^*$$

which is a contradiction to assumption (4) in Definition 4.2. It follows that there exist no objects of $K$ processed after $x_n$. Combining both results yields the desired propositions. □

---

Theorem 5.2 (Approximate FINEX Clusters). *Let $\widetilde{O}_{\varepsilon, MinPts}$ be a FINEX-ordering computed by Algorithm 2. Considering an approximate cluster $S$ w.r.t. $(\varepsilon^*, MinPts)$ according to Definition 4.2, the following propositions hold:*

(a) $\exists$ a cluster $K$ w.r.t. $(\varepsilon^*, MinPts)$ such that $S \subseteq K$.
(b) $\forall p \in K : |N_\varepsilon(p)| \geq MinPts \implies p.P \leq x_n.P$.
(c) $\forall p \in K : |N_{\varepsilon^*}(p)| \geq MinPts \implies p \in S$.

Proof. **(a):** The proof is equal to the one for proposition (a) of Theorem 4.3. **(b) and (c):** The proof is analogous to the one for propositions (b) and (c) of Theorem 4.3. All we have to do is restrict objects $y$ in Lemma A.1 and $x_n$ in Lemma A.2 to be core objects w.r.t. $\varepsilon$. $\qquad\square$

---

Theorem 5.3 (Clustering of Non-Cores). *Let $x \in \widetilde{O}_{\varepsilon, MinPts}$ be a non-core object w.r.t. $(\varepsilon, MinPts)$. Given a distance threshold $\varepsilon^* \leq \varepsilon$, Algorithm 1 applied to $\widetilde{O}_{\varepsilon, MinPts}$ correctly assigns $x$ to*

- *the noise group if $x$ is a noise object w.r.t. $(\varepsilon^*, MinPts)$, or*
- *a corresponding cluster if $x$ is a border object w.r.t. $(\varepsilon^*, MinPts)$.*

Proof. If $x \in \widetilde{O}_{\varepsilon, MinPts}$ is noise w.r.t. $(\varepsilon^*, MinPts)$, then $x$ cannot belong to any approximate cluster $S$ from Theorem 5.2. Thus, the proposition follows.

Now, let $x$ be a border object w.r.t. $(\varepsilon^*, MinPts)$. Hence, there exists an object $o$ with

$$R_{\varepsilon, MinPts}(x, o) \leq \varepsilon^*$$

According to Definition 5.1 it holds that

$$x.R = \min\{R_{\varepsilon, MinPts}(x, o) : o \in D\}$$

Combining both conclusions yields

$$x.R \leq \varepsilon^*$$

It remains to show that $x$ is assigned to one of the clusters it belongs to (and not just to anyone). Let $p$ be the object that last inserted or decreased $x$ in the priority queue. From

$$x.R = R_{\varepsilon, MinPts}(x, p) \leq \varepsilon^*$$

we know that $p$ is a core object w.r.t. $\varepsilon^*$ and $p$ and $x$ belong to the same cluster. From Theorem 5.2 (c) it follows that $p$ will be detected as a cluster member by Algorithm 1. Moreover, it must hold that $p.P < x.P$. Thus, the priority queue was nonempty after $p$ has been processed (at least $x$ was contained) and it holds

$$\forall o \in \widetilde{O}_{\varepsilon, MinPts} \text{ with } p.P < o.P < x.P : o.R \leq x.R \leq \varepsilon^*$$

Consequently, both $p$ and $x$ are assigned to the same cluster by Algorithm 1. $\qquad\square$

---

Theorem 5.4 (Clustering of Former-Cores). *Let $x \in D$ be a former-core w.r.t. $\varepsilon^* \leq \varepsilon$ and $MinPts$, and let $\widetilde{O}_{\varepsilon, MinPts}$ and $O_{\varepsilon, MinPts}$ be the FINEX- and OPTICS-ordering, respectively. For $\varepsilon^*$, it holds that Algorithm 1 with $\widetilde{O}_{\varepsilon, MinPts}$ identifies $x$ as cluster member iff Algorithm 1 with $O_{\varepsilon, MinPts}$ identifies $x$ as cluster member.*

Proof. We assume a stable priority queue that pops its elements in insertion order in case of ties (i.e., when the priority of two elements is equal). Let $x \in D$ be a former core, i.e., a core object w.r.t. $(\varepsilon, MinPts)$ and a border object w.r.t. $(\varepsilon^*, MinPts)$, $\varepsilon^* \leq \varepsilon$. Hence, $\varepsilon^* < x.C \leq \varepsilon$ in both $\widetilde{O}_{\varepsilon, MinPts}$ and $O_{\varepsilon, MinPts}$ (cf. theorems 4.3 and 5.2). If $x$ is identified as a cluster member from $\widetilde{O}_{\varepsilon, MinPts}$, then there exists a core $c \in \widetilde{O}_{\varepsilon, MinPts}$ and a sequence of objects $x_1, \ldots, x_n \in \widetilde{O}_{\varepsilon, MinPts}$ with the following properties

- $c.C \leq \varepsilon^*$ and $c.R > \varepsilon^*$

- $x_1 = c, x_n = x$
- $x_{i+1}.P = x_i.P + 1$ for $i = 1, \ldots, n - 1$
- $x_i.R \leq \varepsilon^*$ for $i = 2, \ldots, n$

In other words, $c$ initializes the approximate cluster in which $x$ is contained. Reprocessing non-core objects w.r.t. $(\varepsilon, MinPts)$ in Algorithm 2 does not affect the reachability distance attribute of core objects w.r.t. $(\varepsilon, MinPts)$. Hence, the priority queue pops core objects in exactly equal order for both FINEX (i.e., Algorithm 2) and OPTICS. Therefore, all core objects of $x_1, \ldots, x_n$ are equally ordered in both $\widetilde{O}_{\varepsilon, MinPts}$ and $O_{\varepsilon, MinPts}$. Moreover, $x_i.R$ for core objects of $x_1, \ldots, x_n$ is equal in both $\widetilde{O}_{\varepsilon, MinPts}$ and $O_{\varepsilon, MinPts}$. In particular, $x.R \leq \varepsilon^*$ also in $O_{\varepsilon, MinPts}$. It follows that there exists an approximate cluster in $O_{\varepsilon, MinPts}$ to which $x$ and $c$ belong. Hence, $x$ will be detected as a cluster member by Algorithm 1 when applied to $O_{\varepsilon, MinPts}$. The other direction of the proof works analogously. □

---

THEOREM 5.6 (EXACT $\varepsilon^*$-QUERY). *Let $\widetilde{O}_{\varepsilon, MinPts}$ be a FINEX-ordering computed by Algorithm 2. Algorithm 1 with $\varepsilon^* \leq \varepsilon$ returns approximate clusters $S_1, \ldots, S_m$ w.r.t. $(\varepsilon^*, MinPts)$. For each $S_i$, we add $o \in \widetilde{O}_{\varepsilon, MinPts}$ to $S_i$ if $o$ is directly density-reachable from some core object of $S_i$ w.r.t. $(\varepsilon^*, MinPts)$ and the following conditions hold:*

1. *$o$ is labeled noise and $\varepsilon^* < o.C \leq \varepsilon$*
2. *$o$ is processed before the first object in $S_i$*
3. *$\exists$ a cluster $K$ w.r.t. $(\varepsilon, MinPts)$ such that $S_i \cup \{o\} \subseteq K$*
4. *$o$ has not yet been added to another $S_j$, $j \neq i$*

*The result is an exact clustering w.r.t. $(\varepsilon^*, MinPts)$ according to Definition 3.5.*

PROOF. This theorem follows from Theorem 5.2 in combination with Theorem 5.3 and Proposition 3.9. □

---

COROLLARY 5.5 (LINEAR EXACT QUERY). *Let $\widetilde{O}_{\varepsilon, MinPts}$ be a FINEX-ordering computed by Algorithm 2. Algorithm 1 with $\varepsilon^* = \varepsilon$ returns an exact clustering w.r.t. $(\varepsilon, MinPts)$ according to Definition 3.5 in linear time (in the number of objects).*

PROOF. There is no object $o$ with $\varepsilon^* < o.C \leq \varepsilon$. Thus, this corollary directly follows from Theorem 5.6. □

---

PROPOSITION 5.7 (MinPts-NESTED CLUSTERS). *Let $MinPts^* \geq MinPts$. For every cluster $K^* \subseteq D$ w.r.t. $(\varepsilon, MinPts^*)$ there exists a cluster $K \subseteq D$ w.r.t. $(\varepsilon, MinPts)$ such that $K^* \subseteq K$.*

PROOF. Let $K^* \subseteq D$ be a cluster w.r.t. $\varepsilon$ and $MinPts^*$. Let $p$ be core object of $K^*$. We have

$$|N_\varepsilon(p)| \geq MinPts^* \geq MinPts$$

By Lemma 1 in [8], the set

$$K := \{x \in D : x \text{ is density-reachable from } p \text{ w.r.t. } \varepsilon \text{ and } MinPts\}$$

is a cluster in $D$ w.r.t. $\varepsilon$ and $MinPts$. Now, let $q \in K^*$ arbitrary. From connectivity in Definition 3.4 it follows that $q$ is density-reachable from $p$ w.r.t. $\varepsilon$ and $MinPts^*$ (recall that $p$ is a core object of $K^*$), i.e., there exist $p_1, \ldots, p_n \in D, p_1 = p$, $p_n = q$ with

$$p_{i+1} \in N_\varepsilon(p_i) \ \wedge \ |N_\varepsilon(p_i)| \geq MinPts^* \ \forall i \in \{1, \ldots, n - 1\}$$

Obviously, it holds that $|N_\varepsilon(p_i)| \geq MinPts$ for all $i$. Consequently, $q$ is density-reachable from $p$ w.r.t. $\varepsilon$ and $MinPts$. It follows that $q \in K$. □